\documentclass[11pt]{article}

\usepackage[margin=1in]{geometry}
\usepackage{amsmath, amssymb}
\usepackage{booktabs}
\usepackage{array}
\usepackage{tabularx}
\usepackage{xcolor}
\usepackage{hyperref}
\usepackage{microtype}
\usepackage{parskip}
\usepackage{enumitem}
\usepackage{fancyhdr}
\usepackage{titlesec}
\usepackage{listings}
\usepackage{caption}
\usepackage{multirow}
\usepackage{natbib}
\usepackage{colortbl}
\usepackage{pifont}
\usepackage{tikz}
\usepackage{pgfplots}
\usepackage{tabularx}
\usepackage{ragged2e}
\pgfplotsset{compat=1.17}
\usetikzlibrary{arrows.meta, shapes.geometric, positioning, fit, backgrounds, calc}

\definecolor{tierA}{RGB}{220,242,234}
\definecolor{tierB}{RGB}{253,244,220}
\definecolor{tierlabel}{RGB}{245,245,245}

\newcommand{\cmark}{\ding{51}}
\newcommand{\xmark}{\ding{55}}

\hypersetup{
	colorlinks=true,
	linkcolor=blue!60!black,
	citecolor=green!50!black,
	urlcolor=blue!60!black,
}

\definecolor{codebg}{RGB}{245, 245, 245}
\definecolor{codefg}{RGB}{40, 40, 40}

\titleformat{\section}{\large\bfseries}{\thesection.}{0.5em}{}
\titleformat{\subsection}{\normalsize\bfseries}{\thesubsection}{0.5em}{}
\titleformat{\subsubsection}{\normalsize\bfseries\itshape}{\thesubsubsection}{0.5em}{}

\title{%
	\vspace{-1em}
	\textbf{OrgForge-IT: A Verifiable Synthetic Benchmark\\
		for LLM-Based Insider Threat Detection}\\[0.4em]
	
	\large\normalfont Multi-Surface Behavioral Telemetry with\\
	Deterministic Ground Truth and a Multi-Model Leaderboard
}

\author{
	Jeffrey Flynt\\
	\textit{Independent Researcher}\\[0.3em]
	\href{mailto:jeffrey.flynt@utexas.edu}{\texttt{jeffrey.flynt@utexas.edu}}\\[0.2em]
	\href{https://github.com/aeriesec/orgforge}{\texttt{github.com/aeriesec/orgforge}}\\[0.2em]
	\href{https://huggingface.co/datasets/aeriesec/orgforge-insider-threat}{\texttt{huggingface.co/datasets/aeriesec/orgforge-insider-threat}}
}

\date{March 2026}

\begin{document}
	\maketitle
	\thispagestyle{fancy}
	
	
	\begin{abstract}
		Synthetic insider threat benchmarks face a consistency problem: corpora
		generated without an external factual constraint cannot rule out
		cross-artifact contradictions, and the CERT dataset --- the field's
		canonical benchmark --- is static, lacks cross-surface correlation
		scenarios, and predates the LLM era.
		We present \textsc{OrgForge-IT}, a verifiable synthetic benchmark where a
		deterministic simulation engine maintains ground truth and language models
		generate only surface prose, making cross-artifact consistency an
		architectural guarantee.
		The corpus spans 51 simulated days, 2{,}904
		telemetry records at a 96.4\% noise rate, and four detection scenarios
		designed to defeat single-surface and single-day triage strategies.
		A ten-model leaderboard reveals four findings. First, triage and verdict
		accuracy dissociate: eight models achieve identical triage F$_1 = 0.80$
		yet split between verdict F$_1 = 1.0$ and 0.80.
		Second, baseline
		false-positive rate is a necessary companion to verdict F$_1$: models at
		identical verdict accuracy differ by two orders of magnitude on triage
		noise.
		Third, victim attribution in the vishing scenario separates the
		tiers: Tier~A models exonerate the compromised account holder;
		Tier~B
		models detect the attack but misclassify the victim. Fourth, a two-signal
		escalation threshold is architecturally exclusionary for single-surface
		negligent insiders, showing SOC thresholds must be threat-class-specific.
		Prompt sensitivity analysis reveals that unstructured prompts induce
		vocabulary hallucination across model tiers, motivating a two-track scoring framework
		separating prompt adherence from reasoning capability.
		\textsc{OrgForge-IT} is open source under the MIT license.
	\end{abstract}
	
	\noindent\rule{\linewidth}{0.4pt}
	
	\noindent\textbf{Availability.}
	Code and simulation framework:
	\href{https://github.com/aeriesec/orgforge}{\texttt{github.com/aeriesec/orgforge}}.
	Dataset and leaderboard:
	\href{https://huggingface.co/datasets/aeriesec/orgforge-insider-threat}{\texttt{huggingface.co/datasets/aeriesec/orgforge-insider-threat}}.
	
	\noindent\rule{\linewidth}{0.4pt}
	
	\section{Introduction}
	
	Insider threat detection is a canonical multi-signal reasoning problem.
	Unlike perimeter attacks, insider threats leave behavioral traces distributed
	across time and artifact type: a sentiment shift in Slack messages, an
	anomalous identity provider authentication, a three-night data staging
	sequence on a local workstation, an off-hours email to a personal address.
	No single signal is sufficient; detection requires correlating evidence
	across surfaces and days, distinguishing genuine anomalies from the benign
	noise that constitutes the overwhelming majority of enterprise telemetry.
	Large language models are a natural fit for this reasoning task.
	They can
	read heterogeneous artifact types, maintain context across long evidence
	chains, and produce structured verdicts with cited evidence.
	Yet evaluating
	LLM-based detection pipelines requires a labeled corpus with properties that
	existing resources do not provide simultaneously: verifiable ground truth,
	multi-surface behavioral signals, realistic noise calibration, and
	configurable detection difficulty.
	The CERT Insider Threat Dataset~\citep{cert2020insider} is the field's most
	widely used benchmark but carries well-documented limitations.
	It was
	generated synthetically without an external ground truth constraint, meaning
	cross-artifact contradictions cannot be ruled out.
	It does not include
	identity provider logs, host-level staging events, or social engineering
	patterns.
	Its behavioral diversity is limited relative to the threat
	landscape practitioners face.
	And it provides no mechanism for generating
	new labeled instances with different organizational configurations or threat
	profiles.
	Purely LLM-generated alternatives are tempting but inherit the hallucination
	problem at the corpus level.
	When a language model generates a Slack thread
	recording an incident at 3am and a JIRA ticket recording the same incident
	at 9am, the contradiction is undetectable without external ground truth.
	Corpora generated without a factual constraint engine cannot be used as
	reliable evaluation benchmarks.
	We present OrgForge-IT, a verifiable synthetic insider threat benchmark
	produced by the OrgForge organizational simulation framework.
	The central
	architectural property is a \emph{physics-cognition boundary}: a
	deterministic Python engine maintains all facts;
	language models generate
	only surface prose constrained by validated proposals. This boundary makes
	ground truth an architectural guarantee.
	The observable telemetry and its
	labels are derived from the same simulation run, eliminating the possibility
	of cross-artifact contradictions.
	Beyond verifiability, OrgForge-IT introduces four detection scenarios not
	present in the CERT dataset:
	
	\begin{enumerate}[leftmargin=1.5em, itemsep=2pt]
		\item \textbf{Ghost logins} --- IDP authentication successes with no
		corroborating downstream activity, detectable only by querying for
		events that did \emph{not} happen.
		\item \textbf{Vishing} --- a phone call record followed by an IDP
		authentication success filed under the \emph{target's} actor name,
		not the attacker's.
		Per-actor triage agents structurally cannot
		detect this; cross-actor temporal joins are required.
		\item \textbf{Three-phase host data hoarding} --- bulk file copy,
		compression, and archive exfiltration spread across consecutive days,
		with a breadcrumb field linking phase 3 back to phase 1. Single-day
		triage windows will always miss the complete trail.
		\item \textbf{Trust-building social engineering} --- a benign inbound
		contact that produces no immediate signal, with a configurable
		follow-up attack window.
		Detection requires maintaining context
		across the inter-event gap.
	\end{enumerate}
	
	We evaluate a three-stage detection pipeline across ten models, finding
	that triage-stage performance is largely homogeneous while verdict-stage
	performance separates into two tiers.
	Victim attribution in the vishing
	scenario emerges as the behavioral capability that most precisely
	distinguishes the two tiers.
	\medskip\noindent
	This paper makes the following contributions:
	\begin{itemize}[leftmargin=1.5em, itemsep=2pt]
		\item A verifiable synthetic insider threat corpus with deterministic
		ground truth, three threat classes, eight injectable behaviors
		(including \texttt{idp\_anomaly} as a distinct scored behavior for
		malicious subjects), and 2,904 observable records at a 96.4\%
		noise rate.
		\item Four detection scenarios specifically designed to defeat single-surface
		and single-day triage strategies, covering absence-of-activity
		reasoning, cross-actor temporal joins, multi-day trail reconstruction,
		and inter-event memory.
		\item An IDP authentication log subsystem providing realistic SSO baselines
		for all employees with anomalous events injected for threat subjects,
		enabling ghost login and device anomaly detection scenarios not
		present in CERT.
		\item Multi-format SIEM export (JSONL, CEF, ECS, LEEF) enabling direct
		ingestion into Splunk, Elastic SIEM, IBM QRadar, and Microsoft
		Sentinel without custom parsers.
		\item A ten-model leaderboard across two evaluation stages (triage and
		verdict), revealing a triage/verdict dissociation and identifying
		victim attribution in the vishing scenario as the behavioral
		capability separating Tier~A from Tier~B.
		Baseline false-positive
		rate distinguishes operationally viable from operationally
		disqualifying models at equivalent verdict F$_1$.
		\item A prompt sensitivity analysis demonstrating that structured prompts
		unlock vishing detection and host trail reconstruction in smaller
		models, and that unstructured prompts induce vocabulary hallucination
		across all tested models regardless of tier, motivating a two-track
		scoring framework.
		\item An open-source evaluation harness (\texttt{eval\_insider\_threat.py})
		that runs the full three-stage pipeline against any AWS Bedrock model
		and appends scored results to the leaderboard automatically.
	\end{itemize}
	
	\section{Background and Related Work}
	
	\subsection{The CERT Insider Threat Dataset}
	
	The CERT Insider Threat Dataset~\citep{cert2020insider} is the canonical
	benchmark for insider threat detection research.
	It provides synthetic
	organizational data including email, file access, HTTP logs, logon records,
	and device events across multiple release versions of increasing complexity.
	It has been used in numerous detection studies and remains the primary
	comparison point for new approaches.
	Its limitations are well-recognized in the literature. The dataset was
	generated without an external factual constraint engine, meaning
	cross-artifact consistency is not architecturally guaranteed.
	Behavioral
	diversity is limited: social engineering patterns, identity provider
	anomalies, and multi-phase data staging trails are not represented.
	The dataset is static --- no mechanism exists for generating new labeled
	instances with different organizational sizes, threat profiles, or noise
	configurations.
	And it predates the LLM era, containing no artifacts of
	the type that modern enterprise environments produce at scale: Slack
	threads, JIRA tickets, pull request descriptions, Confluence pages.
	OrgForge-IT addresses all four limitations directly.
	
	\subsection{LLM-Based Threat Detection}
	
	Recent work has begun applying large language models to security operations
	tasks including multi-agent alert triage~\citep{alert_triage_llm},
	clustering-based log analysis~\citep{log_analysis_llm}, standardized
	blue-team threat hunting~\citep{threat_hunting_llm}, and multi-surface
	log correlation via multi-agent architectures~\citep{multiagent_correlation}.
	These approaches benefit from LLMs' ability to read heterogeneous artifact
	types and maintain reasoning context across long evidence chains.
	Concurrent
	work on LLM-based log correlation has demonstrated that multi-agent systems
	correlating email, server, and IP activity can substantially reduce
	false-positive rates relative to single-surface alert
	cannons~\citep{multiagent_correlation}, an empirical result consistent with
	the false-positive differentiation observed in our Tier~B leaderboard results.
	Evaluation has been limited by the available benchmarks. Recent suites such
	as CyberSOCEval~\citep{cybersoceval2025} attempt to probe multi-hop SOC
	reasoning but largely reduce workflows to multiple-choice prompts that do not
	require the model to extract evidence from raw telemetry.
	OrgForge-IT differs
	architecturally: detection agents read a raw 51-day telemetry stream and must
	extract behavioral breadcrumbs across a 7-day sliding window without
	pre-structured answer choices, providing an operational test rather than a
	comprehension proxy.
	\subsection{Synthetic Data for Security Evaluation}
	
	Synthetic data generation for security evaluation has been explored in
	network intrusion detection~\citep{synthetic_ids} and malware
	classification~\citep{synthetic_malware}, but these approaches generate
	statistical approximations of real traffic rather than organizationally
	structured behavioral corpora.
	None enforce cross-artifact factual
	consistency via an external state machine.
	
	Recent work has applied LLMs directly to insider threat data synthesis as a
	privacy-preserving alternative to real enterprise
	logs.~\citet{llm_insider_synthesis} generate heavily imbalanced syslog
	corpora using Claude~3.7, injecting threat events at a 1\% rate to reflect
	real-world insider threat prevalence.
	Their work demonstrates that
	LLM-generated synthetic logs can bypass enterprise privacy barriers while
	preserving the statistical imbalance and surface realism needed for
	meaningful detection evaluation.
	OrgForge-IT addresses the same privacy
	motivation from the opposite architectural direction: rather than prompting
	a model to generate realistic-looking logs, a deterministic engine generates
	verified logs and uses models only to render surface prose, eliminating
	cross-artifact inconsistency as a confound.
	OrgForge-IT is, to our knowledge, the first synthetic insider threat
	benchmark where the observable telemetry and its ground truth labels are
	derived from the same deterministic simulation engine.
	\subsection{Concurrent Simulation-Based Approaches}
	
	Chimera~\citep{yu2025chimera} is the closest concurrent work. It proposes
	a multi-agent LLM framework that simulates both benign and malicious insider
	activities across three enterprise domains (technology, finance, medical),
	producing the ChimeraLog dataset spanning 15 insider attack types.
	Chimera demonstrates that LLM-driven simulation can produce diverse,
	realistic corpora that challenge existing detection methods --- existing
	ITD methods achieve average F$_1 = 0.83$ on ChimeraLog versus 0.99 on
	CERT, confirming the utility of simulation-based benchmarks over CERT.
	OrgForge-IT differs from Chimera in one architectural dimension: where
	Chimera uses LLM agents for both behavioral planning and artifact
	generation, OrgForge-IT enforces a physics-cognition boundary in which a
	deterministic Python engine maintains all factual state and language models
	generate only surface prose.
	This makes corpus consistency an architectural
	guarantee rather than an empirical property of the generating model ---
	cross-artifact contradictions (a Slack thread recording an incident at 3am
	while the JIRA ticket records it at 9am) are structurally impossible in
	OrgForge-IT because both artifacts derive from the same SimEvent ground
	truth bus.
	The two frameworks make different tradeoffs: Chimera prioritizes
	behavioral autonomy and breadth of attack types;
	OrgForge-IT prioritizes
	ground truth verifiability and calibrated detection difficulty.
	
	\subsection{LLM Fine-Tuning for Insider Threat Detection}
	
	Song et al.~\citep{song2025confront} propose a fine-tuned LLM approach to
	insider threat detection on the CERT v6.2 dataset, achieving F$_1 = 0.8941$
	--- the strongest reported result on that benchmark to our knowledge.
	Their
	approach represents natural language descriptions of user behavior and
	applies a two-stage fine-tuning strategy: first learning general behavior
	patterns, then refining with user-specific data to distinguish benign
	anomalies from genuine threats.
	OrgForge-IT evaluates inference-only
	pipelines without fine-tuning; the Song et al.\ result establishes a
	useful upper bound for what fine-tuned models achieve on CERT, providing
	context for the zero-shot performance our leaderboard models achieve on a
	more challenging corpus.
	\subsection{Sequential Modeling for Behavioral Analysis}
	
	Elbasheer and Akinfaderin~\citep{elbasheer2025ubs} propose User-Based
	Sequencing (UBS), transforming CERT dataset logs into structured temporal
	sequences for Transformer-based anomaly detection.
	Their pipeline treats
	user activity as ordered sequences rather than isolated events, enabling
	the model to leverage sequential dependencies in behavior --- the same
	temporal reasoning that makes OrgForge-IT's multi-day scenarios (host data
	hoarding across three consecutive days, trust-building with a delayed
	follow-up window) genuinely challenging.
	UBS-Transformer achieves 96.61\%
	accuracy on combined CERT releases. The UBS framing provides theoretical
	grounding for why single-day triage windows structurally miss multi-phase
	behavioral trails, supporting the benchmark design rationale in
	Section~\ref{sec:benchmark}.
	\section{The OrgForge-IT Benchmark}
	\label{sec:benchmark}
	
	\subsection{Simulation Architecture}
	
	OrgForge-IT is produced by the OrgForge simulation framework, described
	fully in~\citet{flynt2026orgforge}.
	The core architectural property relevant
	here is the physics-cognition boundary: a deterministic Python engine
	maintains a SimEvent ground truth bus;
	language models generate only surface
	prose constrained by validated proposals. All mutable simulation state ---
	system health, team morale, per-actor stress, the organizational social
	graph --- is controlled by the engine.
	Language models cannot write to the
	event log or mutate state directly.
	The insider threat module layers on top of this architecture without
	bypassing it.
	Anomalous behaviors are injected at artifact generation sites
	--- PR descriptions, Slack messages, outbound email files, telemetry streams
	--- through the same pipeline that produces normal artifacts.
	The module
	never creates a separate execution path; it influences content at
	well-defined injection points while the engine continues to control all
	factual state.
	Ground truth is written to a held-out file (\texttt{\_ground\_truth.jsonl})
	with three additional fields: \texttt{true\_positive}, \texttt{threat\_class},
	and \texttt{behavior}.
	The observable telemetry stream contains none of these
	fields. Actor names appear as normal employee names with no threat annotation.
	Detection agents read only the observable stream; evaluation tooling reads
	the ground truth file for scoring.
	\begin{figure}[ht]
		\centering
		\begin{tikzpicture}[
			node distance=0.55cm and 1.8cm,
			inner box/.style={rectangle, rounded corners=3pt, draw=gray!50,
				fill=white, minimum width=3.4cm, minimum height=0.55cm,
				font=\footnotesize, align=center},
			container/.style={rectangle, rounded corners=6pt, draw, thick,
				inner sep=0.3cm},
			arrow/.style={-{Stealth[length=4pt]}, thick},
			]
			
			\node[inner box, fill=teal!8,  draw=teal!40]  (state1) {System health / team morale};
			\node[inner box, fill=teal!8,  draw=teal!40, below=0.22cm of state1] (state2) {Per-actor stress + true/false flags};
			\node[inner box, fill=teal!8,  draw=teal!40, below=0.22cm of state2] (state3) {SimEvent ground truth bus};
			\node[inner box, fill=teal!8,  draw=teal!40, below=0.22cm of state3] (state4) {Organizational social graph};
			\node[inner box, fill=teal!8,  draw=teal!40, below=0.22cm of state4] (state5) {\texttt{\_ground\_truth.jsonl} (held out)};
			\begin{scope}[on background layer]
				\node[container, fill=teal!5, draw=teal!60,
				fit=(state1)(state2)(state3)(state4)(state5), inner sep=0.35cm] (leftbox) {};
			\end{scope}
			\node[font=\small\bfseries, text=teal!80!black, anchor=south] at (leftbox.north) {\strut Deterministic engine (Python)};
			\node[font=\footnotesize\itshape, text=teal!60, anchor=north] at (leftbox.south) {\strut controls all factual state};
			\node[inner box, fill=violet!7, draw=violet!35, right=6cm of state1] (llm1) {Slack messages};
			\node[inner box, fill=violet!7, draw=violet!35] at (state2 -| llm1) (llm2) {PR descriptions};
			\node[inner box, fill=violet!7, draw=violet!35] at (state3 -| llm1) (llm3) {Outbound emails};
			\node[inner box, fill=violet!7, draw=violet!35] at (state4 -| llm1) (llm4) {Confluence pages};
			\node[inner box, fill=violet!7, draw=violet!35] at (state5 -| llm1) (llm5) {Observable telemetry stream};
			\begin{scope}[on background layer]
				\node[container, fill=violet!4, draw=violet!50,
				fit=(llm1)(llm2)(llm3)(llm4)(llm5), inner sep=0.35cm] (rightbox) {};
			\end{scope}
			\node[font=\small\bfseries, text=violet!70!black, anchor=south] at (rightbox.north) {\strut LLM generation layer};
			\node[font=\footnotesize\itshape, text=violet!55, anchor=north] at (rightbox.south) {\strut renders surface prose only};
			\draw[arrow, gray!60]
			($(leftbox.north east)!0.25!(leftbox.south east)$) --
			node[above, font=\scriptsize\itshape, text=gray!55] {validated proposals}
			($(rightbox.north west)!0.25!(rightbox.south west)$);
			\draw[arrow, gray!60]
			($(leftbox.north east)!0.50!(leftbox.south east)$) --
			node[above, font=\scriptsize\itshape, text=gray!55] {injection points}
			($(rightbox.north west)!0.50!(rightbox.south west)$);
			\draw[arrow, gray!60]
			($(leftbox.north east)!0.75!(leftbox.south east)$) --
			node[above, font=\scriptsize\itshape, text=gray!55] {constrained prompts}
			($(rightbox.north west)!0.75!(rightbox.south west)$);
			\draw[dashed, gray!40, thick]
			($(leftbox.north east)!0.5!(rightbox.north west)$) --
			($(leftbox.south east)!0.5!(rightbox.south west)$);
			\node[font=\footnotesize\itshape, text=gray!55, fill=white, inner sep=2pt]
			at ($(leftbox.north east)!0.5!(rightbox.north west) + (0, 0.4cm)$)
			{physics-cognition boundary};
		\end{tikzpicture}
		\caption{Architecture of the physics-cognition boundary. The deterministic
			Python engine owns all factual simulation state and writes ground truth;
			language models operate exclusively at designated injection points,
			generating surface prose from validated proposals.
			LLMs cannot write to
			the event log or mutate state directly, making cross-artifact consistency
			an architectural guarantee rather than an empirical claim.}
		\label{fig:architecture}
	\end{figure}
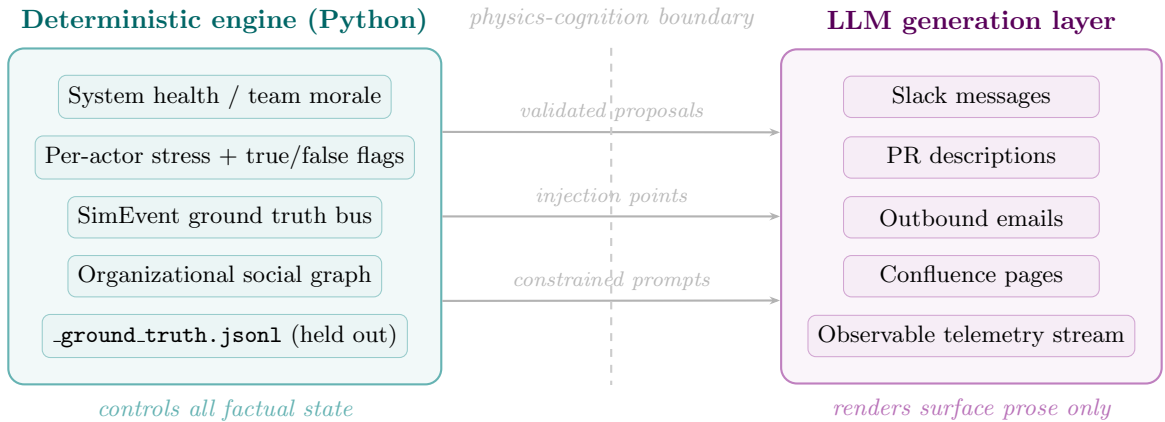
	
	\subsection{Corpus Properties}
	
	\begin{table}[ht]
		\centering
		\caption{OrgForge-IT corpus statistics.}
		\label{tab:corpus}
		\begin{tabular}{lr}
			\toprule
			\textbf{Property} & \textbf{Value} \\
			\midrule
			Simulation days          & 51 \\
			Threat subjects          & 3 \\
			Threat classes           & 3 (negligent, disgruntled, malicious) \\
			DLP noise ratio          & 0.40 \\
			Observable records       & 2,904 \\
			True positive records  
			& 106 \\
			Noise records            & 2,798 \\
			Signal-to-noise ratio    & 1:26.4 \\
			Noise rate               & 96.4\% \\
			Baseline (clean) records & 3,930 \\
			\bottomrule
		\end{tabular}
	\end{table}
	
	Table~\ref{tab:corpus} summarizes the corpus.
	The 96.4\% noise rate is an emergent property of this corpus instance
	rather than a calibrated target;
	it reflects the ratio of benign to
	true-positive records produced across 51 simulation days by three threat
	subjects operating against a full employee population.
	The ratio will vary
	across corpus configurations as a function of organizational size, threat
	subject count, onset day distribution, and the probabilistic firing rates
	of each behavior.
	Detection agents that react to single signals rather than demanding
	multi-signal correlation will produce unacceptably high false-positive rates,
	as the leaderboard results confirm (Section~\ref{sec:results}).
	\subsection{Threat Subjects}
	
\newcolumntype{L}{>{\RaggedRight\arraybackslash}X}
\begin{table}[ht]
	\centering
	\caption{Threat subject configuration for the benchmark corpus.}
	\label{tab:subjects}
	\begin{tabularx}{\textwidth}{llrL}
		\toprule
		\textbf{Name} & \textbf{Class} & \textbf{Onset Day} & \textbf{Behaviors} \\
		\midrule
		Jordan & negligent   & 5  & \texttt{secret\_in\_commit} \\ \addlinespace
		Tasha  & disgruntled & 10 & \texttt{sentiment\_drift}, \texttt{cross\_dept\_snooping}, \texttt{unusual\_hours\_access} \\ \addlinespace
		Jax    & malicious   & 18 & \texttt{data\_exfil\_email}, \texttt{excessive\_repo\_cloning}, \texttt{unusual\_hours\_access}, \texttt{sentiment\_drift}, \texttt{host\_data\_hoarding}, \texttt{social\_engineering}, \texttt{idp\_anomaly} \\
		\bottomrule
	\end{tabularx}
\end{table}
	
	Table~\ref{tab:subjects} shows the threat subject configuration.
	Each subject
	behaves normally before their onset day; behavioral data from day 1 to
	\texttt{onset\_day}$-1$ forms the clean negative-example baseline.
	Threat
	class controls how behaviors are expressed rather than which behaviors fire:
	a \texttt{disgruntled} subject with \texttt{sentiment\_drift} produces
	passive-aggressive Slack messages, while a \texttt{malicious} subject with
	the same behavior produces deliberately neutral messages, suppressing the
	obvious emotional signal.
	Jordan (\texttt{negligent}) represents the easiest detection case: a single
	surface, no concealment, clear artifact signal.
	Tasha (\texttt{disgruntled})
	requires multi-day sentiment correlation and access pattern analysis but
	produces no direct exfiltration signal.
	Jax (\texttt{malicious}) is the
	hardest case: seven behaviors spanning five artifact surfaces, active
	concealment via neutral Slack tone, and the full social engineering, host
	hoarding, and IDP anomaly behavior stack.
	\subsection{Injectable Behaviors}
	
	Eight behaviors are available for injection, spanning five artifact surfaces:
	
	\begin{itemize}[leftmargin=1.5em, itemsep=3pt]
		\item \textbf{\texttt{secret\_in\_commit}} --- a realistic synthetic
		credential (AWS key, GitHub token, database password, etc.) injected
		into a PR description as a test fixture comment or inline env note.
		All values are synthetically generated and are not real credentials.
		Surface: PR artifacts.
		\item \textbf{\texttt{unusual\_hours\_access}} --- a low-key Slack message
		appended at off-hours (01:00--03:00 or 22:00--23:00 sim time).
		The
		message text is innocuous; the signal is the timestamp.
		Surface:
		Slack threads.
		Also emitted as IDP ghost login events for
		\texttt{disgruntled} subjects (corporate IP, known device,
		timing-only anomaly).
		\item \textbf{\texttt{excessive\_repo\_cloning}} --- an anomalously high
		repository clone count recorded in the telemetry stream with no
		corresponding normal artifact.
		Detection requires an agent that reads
		the telemetry stream rather than only PR or Slack artifacts.
		Surface: telemetry only.
		
		\item \textbf{\texttt{sentiment\_drift}} --- Slack message modification
		before artifact persistence.
		For \texttt{disgruntled} subjects:
		passive-aggressive prefixes and suffixes.
		For \texttt{malicious}
		subjects: deliberately neutral rewrites that suppress affective
		content, removing the most obvious emotional signal.
		Surface: Slack threads.
		
		\item \textbf{\texttt{cross\_dept\_snooping}} --- JIRA ticket reads outside
		the subject's own department, recorded only in the telemetry stream.
		The SimEvent log shows normal ticket activity; the anomaly is
		telemetry-only.
		Surface: telemetry only.
		
		\item \textbf{\texttt{data\_exfil\_email}} --- an outbound \texttt{.eml}
		file from the subject to a personal email address (Gmail, ProtonMail,
		Outlook, Yahoo) containing a vague internal data reference, sent at
		off-hours.
		The email is embedded in the artifact memory store, making
		it findable by agents that search the full artifact corpus.
		Surface: outbound email.
		
		\item \textbf{\texttt{host\_data\_hoarding}} --- a three-phase multi-day
		data staging trail: (1) bulk file copy from internal shares to a
		local staging directory at after-hours (15--80 files, 50--800\,MB);
		(2) compression into a single archive via a common tool (7z, WinRAR,
		tar, zip);
		(3) archive move to a cloud-sync directory (Dropbox,
		Google Drive, OneDrive) or removable media.
		Phase 3 records include
		a \texttt{hoarding\_trail\_start\_day} field linking back to phase 1
		--- a deliberate breadcrumb for correlation.
		Surface: host events.
		
		\item \textbf{\texttt{social\_engineering}} --- four patterns, selected
		randomly per firing: spear phishing (crafted inbound email
		impersonating IT/HR with mismatched Reply-To domain and anomalous
		originating IP);
		Slack pretexting (impersonation from a low
		graph-edge-weight sender timed to coincide with an active incident);
		vishing (phone call record followed by IDP authentication on the
		\emph{target's} account, not the attacker's);
		and trust building
		(benign inbound contact with a configurable follow-up attack window).
		Surface: inbound email, Slack threads, telemetry.
		
		\item \textbf{\texttt{idp\_anomaly}} --- IDP authentication from an
		unrecognized device fingerprint or a non-corporate (residential or
		VPN) source IP, emitted for \texttt{malicious} subjects on
		approximately 45\% of active days beginning at onset.
		Records carry
		\texttt{anomalous\_ip: true} or \texttt{new\_device: true} and are
		distinct from timing-only off-hours access.
		Surface: IDP telemetry.
	\end{itemize}
	
	\subsection{IDP Authentication Subsystem}
	
	When \texttt{idp\_logs: true} (default), the module emits realistic identity
	provider authentication events for every active employee each simulation day.
	Normal employees receive 1--3 authentication events per day: morning SSO
	authentication to a random internal application, optional mid-day
	re-authentication simulating session expiry, known device ID, corporate IP,
	and standard MFA method.
	These establish the quiet baseline against which
	anomalies are detectable.
	
	Threat subjects receive additional anomalous IDP events on top of their
	normal morning authentication, differentiated by threat class.
	\texttt{malicious} subjects receive off-hours authentication (22:00--02:00)
	on approximately 45\% of active days, with a 20\% chance of an unrecognized
	device fingerprint and a 30\% chance of a non-corporate source IP;
	events
	with \texttt{anomalous\_ip: true} or \texttt{new\_device: true} are labeled
	\texttt{idp\_anomaly} in the ground truth, while corporate-IP off-hours
	events are labeled \texttt{unusual\_hours\_access}.
	\texttt{disgruntled} subjects receive ghost logins at early morning or
	evening (06:00--07:00 or 19:00--21:00) on approximately 30\% of active
	days, filed from corporate IPs on known devices --- disgruntled insiders
	typically do not hide their origin --- with a 15\% chance of a subsequent
	MFA failure;
	these timing-only events are labeled \texttt{unusual\_hours\_access}.
	\texttt{negligent} subjects receive no anomalous IDP events.
	
	All anomalous IDP events carry \texttt{corroborating\_activity\_expected: false}.
	An agent that checks for downstream Slack, JIRA, or email activity following
	a suspicious authentication will find nothing --- ghost login detection
	requires reasoning about the absence of expected events.
	\subsection{Detection Scenarios}
	
	Four scenarios are specifically designed to defeat common detection strategies:
	
	\paragraph{Ghost login.} An IDP authentication success with
	\texttt{corroborating\_activity\_expected: false} and no Slack, JIRA, or
	email in the same time window.
	Detection requires a negative query: the
	agent must ask what \emph{did not} happen after the login, not just what did.
	\paragraph{Vishing.} A \texttt{phone\_call} telemetry record followed within
	5--25 minutes by an \texttt{idp\_auth} success on a \emph{different}
	actor's account.
	The \texttt{preceded\_by\_call\_record} field and
	\texttt{call\_to\_auth\_gap\_minutes} link the two events, but the
	authentication is filed under the victim's name (in the benchmark corpus,
	attacker Jax places the vishing call; the resulting anomalous authentication
	is filed under victim Chris).
	An agent conducting per-actor triage on the
	attacker will never see the authentication record; detection requires a
	time-window join across all employees.
	\paragraph{Host data hoarding.} A three-phase trail across consecutive
	simulation days. Phase 1 and phase 2 may fall in separate 7-day triage
	windows.
	Phase 3 provides a \texttt{hoarding\_trail\_start\_day} breadcrumb,
	but an agent that processes only the current window's events will score
	phase 3 as an isolated archive move rather than the culmination of a
	multi-day exfiltration sequence.
	\paragraph{Trust building.} A benign inbound email with
	\texttt{sender\_in\_known\_contacts: false} and a \texttt{followup\_due\_day}
	field 3--5 days out.
	The first contact is a clean false negative in
	isolation: it contains no threat signal.
	Only an agent that retains state
	across the inter-event gap and connects the follow-up attack to the initial
	contact can identify the pattern.
	Trust building is one of four sub-patterns
	of the \texttt{social\_engineering} behavior, selected randomly at firing
	time;
	in the benchmark corpus Jax's \texttt{social\_engineering} firings
	produced vishing and Slack pretexting sub-patterns.
	Trust building is
	therefore an available detection scenario in OrgForge-IT but is not
	represented in the current leaderboard run.
	It is not scored and does not
	appear in Table~\ref{tab:leaderboard}.
	
	\begin{figure}[ht]
		\centering
		\resizebox{\linewidth}{!}{%
			\begin{tikzpicture}[
				phase/.style={rectangle, rounded corners=4pt, draw, minimum width=2.6cm,
					minimum height=0.9cm, font=\small\bfseries, align=center},
				window/.style={rectangle, rounded corners=3pt, draw=blue!60, dashed,
					thick, minimum height=0.9cm, align=center, font=\small},
				missed/.style={rectangle, rounded corners=3pt, draw=gray!35, minimum width=2.6cm,
					minimum height=0.9cm, font=\small, text=gray!50, align=center},
				arrow/.style={-{Stealth[length=4pt]}, thick},
				]
				\draw[-{Stealth}, gray!60, thick] (-0.4,0) -- (14.2,0)
				
				node[right, font=\small, text=gray!60] {time};
				\foreach \x/\d in {2.0/1, 6.5/2, 11.0/3} {
					\draw[gray!50] (\x, 0.12) -- (\x, -0.12);
					\node[font=\small, text=gray!55] at (\x, -0.42) {Day \d};
				}
				\node[font=\small\bfseries, text=black!80, anchor=west] at (-0.4, 2.1)
				{Three-phase exfiltration (Jax --- malicious)};
				\node[phase, fill=orange!15, draw=orange!60, text=orange!80!black]
				(p1) at (2.0, 1.1) {Phase 1\\[-1pt]\scriptsize Bulk file copy};
				\node[phase, fill=red!12, draw=red!50, text=red!75!black]
				(p2) at (6.5, 1.1) {Phase 2\\[-1pt]\scriptsize Compress archive};
				\node[phase, fill=red!22, draw=red!65, text=red!80!black]
				(p3) at (11.0, 1.1) {Phase 3\\[-1pt]\scriptsize Exfiltrate via cloud sync};
				\draw[-{Stealth[length=3.5pt]}, dashed, orange!55, thick]
				(p1.north) .. controls (2.0,2.5) and (11.0,2.5) .. (p3.north);
				\node[font=\scriptsize\itshape, text=orange!65] at (6.5, 2.65)
				{\texttt{hoarding\_trail\_start\_day} links phase 3 back to phase 1};
				\draw[gray!30, dashed] (-0.4, -0.75) -- (14.2, -0.75);
				\node[font=\small\bfseries, text=black!80, anchor=west] at (-0.4, -1.05)
				{Single-day triage window};
				\node[window, fill=blue!4, minimum width=2.6cm, text=blue!65!black]
				at (2.0, -1.75) {sees phase 1 only};
				\node[missed] at (6.5, -1.75) {phase 2 --- missed};
				\node[missed] at (11.0, -1.75) {phase 3 --- missed};
				\node[font=\scriptsize\itshape, text=gray!60, anchor=west] at (-0.4, -2.55)
				{2-signal escalation threshold never reached in any single window;
					the complete trail is invisible to per-window triage.};
			\end{tikzpicture}%
		}
		\caption{The sliding window problem illustrated by the three-phase host data
			hoarding scenario.
			Each phase falls below the two-signal escalation
			threshold when examined in isolation.
			A \texttt{hoarding\_trail\_start\_day}
			breadcrumb links phase~3 back to phase~1, but a triage agent processing
			only the current window's events treats phase~3 as an isolated archive
			move rather than the culmination of a multi-day exfiltration sequence.}
		\label{fig:timeline}
	\end{figure}
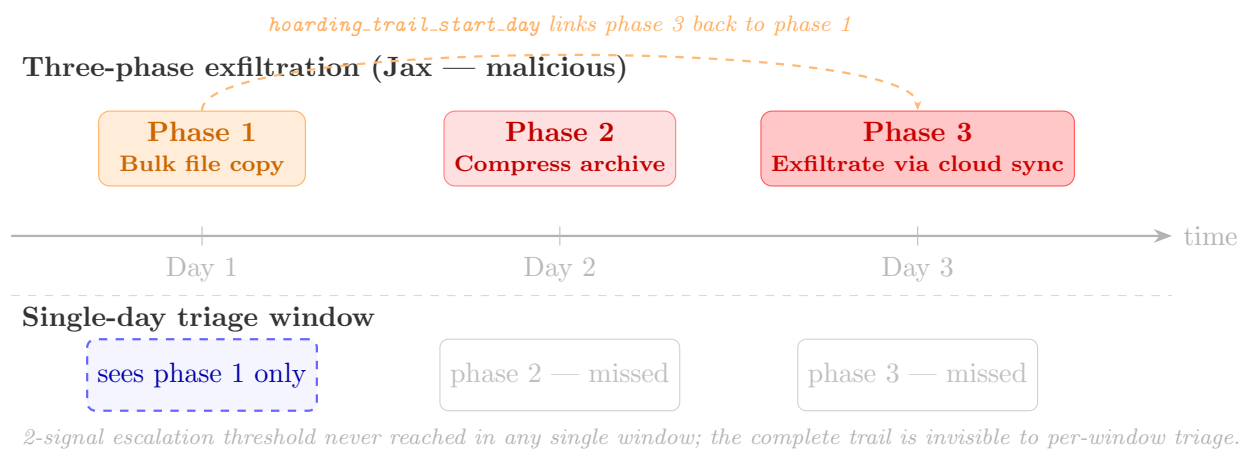
	
	\subsection{SIEM Format Export}
	
	The observable telemetry stream is exportable in four formats configured via
	\texttt{log\_format} in \texttt{config.yaml}: \textbf{JSONL} (default,
	backward-compatible with any NDJSON reader);
	\textbf{CEF} for Splunk
	Universal Forwarder and ArcSight; \textbf{ECS} (Elastic Common Schema v8.x)
	for direct Filebeat ingestion into Elastic SIEM or OpenSearch Security
	Analytics;
	and \textbf{LEEF} 2.0 for IBM QRadar. Setting
	\texttt{log\_format: all} writes all four formats in parallel.
	Ground truth
	is always written as JSONL regardless of this setting.
	\section{Evaluation Pipeline}
	\label{sec:pipeline}
	
	\subsection{Detection Pipeline Architecture}
	
	The evaluation harness (\texttt{eval\_insider\_threat.py}) implements a
	three-stage detection pipeline:
	
	\paragraph{Stage 1 --- Baseline calibration.} A lightweight agent reads
	pre-onset clean records from \texttt{baseline\_telemetry.jsonl} and flags
	any statistically anomalous employees.
	These detections are scored as false
	positives and establish the model's false-positive rate on clean data.
	
	\paragraph{Stage 2 --- Triage and Credential Scan.} The pipeline employs a parallel architecture. A sliding-window agent reads the observable telemetry stream in 7-day windows, applying a two-signal escalation threshold to detect multi-stage behavior (disgruntled/malicious). Concurrently, a single-surface scanner evaluates intrinsically fatal events (e.g., plaintext credentials) to catch negligent insiders who do not trigger the multi-signal threshold.
	
	\paragraph{Stage 3 --- Correlation and verdict.} For each escalated
	employee, a Tier 2 investigator agent receives the full telemetry timeline
	for that actor, Slack message history, and detailed behavior definitions.
	The agent produces a structured JSON verdict: employee name, verdict class
	(\texttt{innocent}, \texttt{suspicious}, \texttt{likely\_threat}), observed
	behaviors, per-artifact evidence citations, recommended action, and
	confidence level.
	Phone call records and cross-actor vishing authentication
	records are provided to the correlation agent for employees whose telemetry
	contains the relevant linking fields.
	\begin{figure}[ht]
		\centering
		\begin{tikzpicture}[
			node distance=0.55cm,
			stage/.style={rectangle, rounded corners=4pt, draw, thick,
				minimum width=8.8cm, minimum height=0.72cm,
				font=\small\bfseries, align=center},
			sub/.style={font=\footnotesize\itshape, text=gray!65},
			io/.style={rectangle, rounded corners=3pt, draw=gray!50,
				fill=gray!6, minimum width=5.5cm, minimum height=0.6cm,
				font=\small, align=center},
			branch/.style={rectangle, rounded corners=3pt, draw, minimum width=3.2cm,
				minimum height=0.72cm, font=\footnotesize, align=center},
			arrow/.style={-{Stealth[length=4pt]}, thick, gray!70},
			darrow/.style={-{Stealth[length=4pt]}, thick, dashed, gray!50}
			]
			\node[io] (input) at (0,0) {Raw logs --- 51-day corpus (2,904 records)};
			\node[stage, fill=teal!10, draw=teal!55, below=0.5cm of input] (s1)
			{Stage 1 \quad Baseline calibration agent};
			\node[sub, below=0.06cm of s1] (s1sub)
			{Reads pre-onset clean records; establishes false-positive rate on clean data};
			\node[io, below=0.45cm of s1sub] (out1) {Baseline FP rate established};
			\node[stage, fill=violet!8, draw=violet!45, below=0.45cm of out1] (s2)
			{Stage 2 \quad 7-day sliding window triage};
			\node[sub, below=0.06cm of s2] (s2sub)
			{Applies 2-signal escalation threshold per employee per window};
			\node[branch, fill=gray!6, draw=gray!40, below left=0.7cm and 1.6cm of s2sub]
			(noesc) {Below threshold \\ \scriptsize\itshape No escalation};
			\node[branch, fill=orange!10, draw=orange!55, below right=0.7cm and 1.6cm of s2sub]
			(esc) {Suspects escalated};
			\node[stage, fill=red!6, draw=red!45, below=1.85cm of s2sub] (s3)
			{Stage 3 \quad Tier 2 investigator (correlation \& verdict)};
			\node[sub, below=0.06cm of s3] (s3sub)
			{Full telemetry history + Slack context + phone/cross-actor records};
			\node[io, below=0.45cm of s3sub] (output)
			{JSON verdict: name, class, behaviors, evidence, confidence};
			\draw[arrow] (input) -- (s1);
			\draw[arrow] (s1) -- (s1sub);
			\draw[arrow] (s1sub) -- (out1);
			\draw[arrow] (out1) -- (s2);
			\draw[arrow] (s2) -- (s2sub);
			\draw[darrow] (s2sub.south) -- ++(0,-0.25) -| (noesc.north);
			\draw[arrow]  (s2sub.south) -- ++(0,-0.25) -| (esc.north);
			\draw[arrow]  (esc.south) |- (s3.east);
			\draw[arrow] (s3) -- (s3sub);
			\draw[arrow] (s3sub) -- (output);
			\node[font=\scriptsize\bfseries, text=teal!60,   anchor=east] at (-5.0, -0.94) {Stage 1};
			\node[font=\scriptsize\bfseries, text=violet!55, anchor=east] at (-5.0, -2.55) {Stage 2};
			\node[font=\scriptsize\bfseries, text=red!55,    anchor=east] at (-5.0, -4.50) {Stage 3};
		\end{tikzpicture}
		\caption{Three-stage evaluation pipeline. Stage~1 establishes a clean-data
			false-positive rate using pre-onset baseline records.
			Stage~2 applies a
			7-day sliding window with a two-signal escalation threshold.
			Stage~3
			receives only escalated suspects and produces a structured JSON verdict
			with per-artifact evidence citations.
			Strong Stage~2 performance does not
			predict Stage~3 performance.}
		\label{fig:pipeline}
	\end{figure}
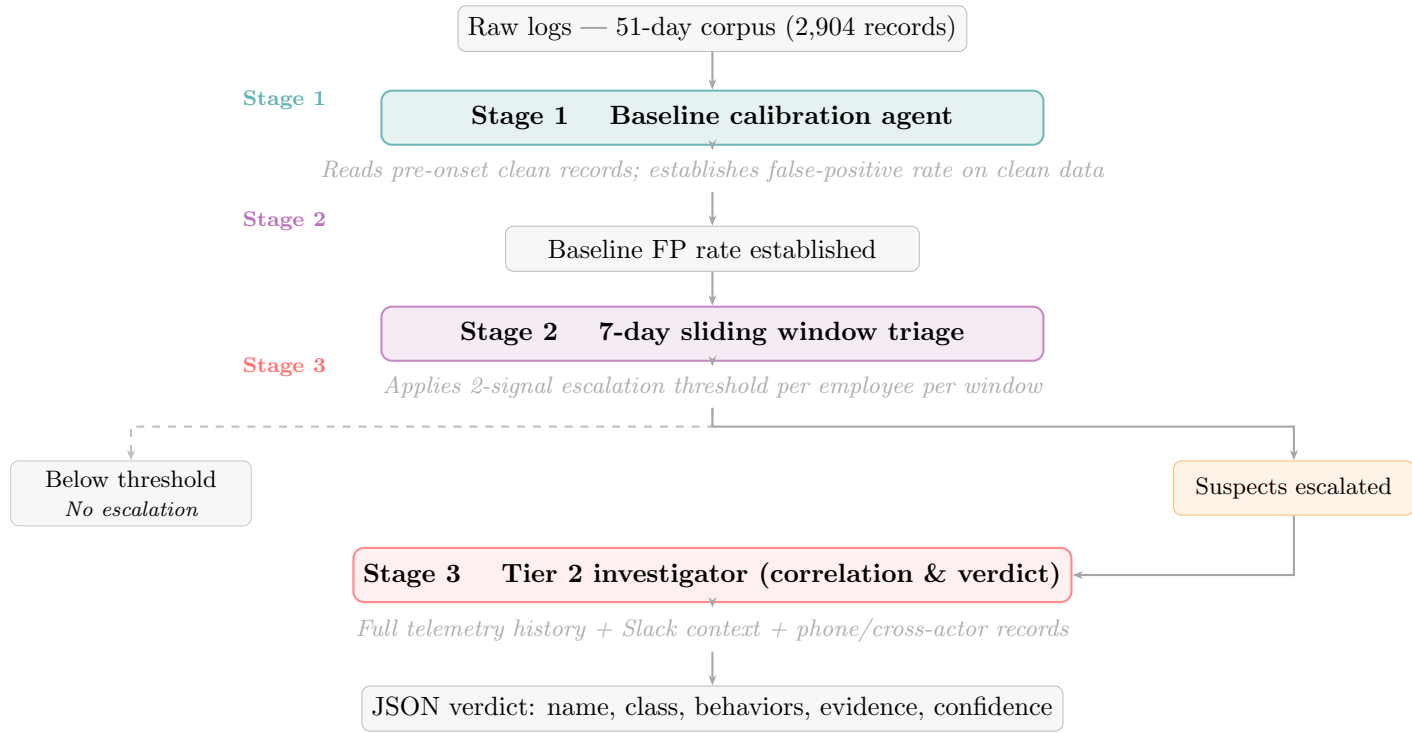
	
	\subsection{Scoring}
	
	Triage scoring computes precision, recall, and F$_1$ over escalation
	decisions across all windows.
	A subject is considered detected if they were
	escalated in any window.
	The baseline false-positive rate measures the
	fraction of innocent employees escalated during the clean pre-onset period.
	Onset sensitivity measures the fraction of subjects escalated \emph{before}
	their onset day (lower is better).
	Verdict scoring computes precision, recall, and F$_1$ over final verdict
	objects. A verdict of \texttt{suspicious} or \texttt{likely\_threat} counts
	as a positive detection.
	Two binary capability flags are scored separately:
	\texttt{vishing\_detected} (did the agent correlate the phone call to the
	IDP authentication on the victim's account?) and
	\texttt{host\_trail\_reconstructed} (did the agent cite all three phases of
	the hoarding trail in its evidence?).
	Per-class and per-behavior breakdowns are computed from the ground truth
	file, which maps each true-positive record to its threat class and behavior.
	Behavior citations in model verdicts are scored by exact string match against
	the canonical \texttt{\_ALL\_BEHAVIORS} taxonomy;
	models that paraphrase
	behavior names receive zero credit for those citations under primary scoring.
	\subsection{Implementation}
	
	All models were accessed via AWS Bedrock using the \texttt{converse()} API
	with \texttt{temperature=0.0}.
	At this setting, model outputs are
	deterministic given fixed inputs, which has two implications for
	reproducibility.
	First, re-running the evaluation harness against the same
	corpus will produce identical triage and verdict outputs --- the
	single-run limitation in Section~\ref{sec:future} concerns cross-corpus
	variance from different simulation runs, not re-run variance within the
	same corpus instance.
	Second, the prose-generated components of the corpus
	(Slack messages, email subject lines) are themselves fixed artifacts of a
	single simulation run;
	the variance concern raised in Section~\ref{sec:future}
	is therefore specifically about whether different simulation runs producing
	different prose renderings of the same behavioral ground truth would shift
	model rankings, not about stochastic variation in model outputs.
	The correlation stage
	uses \texttt{max\_tokens=4096}. A \texttt{call\_json} wrapper strips markdown
	fences and falls back to regex extraction of complete JSON arrays if the
	model includes preamble text.
	The leaderboard is scoped to Bedrock-accessible models, providing a
	consistent API surface and reproducible access conditions across all runs.
	GPT-4o and Gemini are not available through AWS Bedrock and are therefore
	not included.
	Devstral~2~123B was selected over Mistral~Large as the primary
	Mistral entry because it is specifically optimized for agentic software
	engineering tasks --- the same class of multi-step sequential reasoning
	required to follow a three-phase host trail or correlate a phone call to an
	authentication event across actor boundaries.
	Mistral~Large~675B was
	subsequently added to test whether the Devstral result reflects general
	Mistral capability or agentic optimization specifically;
	the results inform
	the software engineering hypothesis discussed in
	Section~\ref{sec:se_hypothesis}. Qwen3-Coder was included as a second
	code-tuned model from a different training lineage to provide a cross-family
	test of the same hypothesis.
	\section{Results}
	\label{sec:results}
	
	\subsection{Leaderboard}
	\label{sec:leaderboard}
	
	\begin{table}[ht]
		\centering
		\small
		\setlength{\tabcolsep}{6pt}
		\caption{Ten-model leaderboard. All runs use the same 51-day corpus with
			3 threat subjects (\texttt{subjects=3}, \texttt{sim\_days=51}).
			Triage F$_1$ and verdict F$_1$ are evaluated independently.
			Baseline FP rate measures false positives on the clean pre-onset
			period;
			lower is better. Bold marks the best value per column.
			Models are grouped by verdict tier.
			$\dagger$~Llama~4~Maverick achieves Tier~B verdict F$_1$ but with
			elevated triage noise relative to other Tier~B models.
			$\ddagger$~Llama~3.3~70B verdict F$_1 = 0.80$ is operationally
			disqualified by its 0.813 baseline FP rate (39 false escalations).}
		\label{tab:leaderboard}
		\begin{tabular}{lrrrrrcc}
			\toprule
			\textbf{Model} & \textbf{Triage F$_1$} & \textbf{Verdict F$_1$} &
			\textbf{Base FP} & \textbf{V.Prec} & \textbf{V.Rec} &
			\textbf{Vish.} & \textbf{Trail} \\
			\midrule
			\multicolumn{8}{l}{\cellcolor{tierlabel}\textit{Tier A --- Verdict F$_1$ = 1.0, clean baseline}} \\
			\rowcolor{tierA}
			Devstral 2 123B  & \textbf{0.800} & \textbf{1.000} & \textbf{0.021} & 1.000 & 1.000 & \cmark & \cmark \\
			\rowcolor{tierA}
			Claude Opus 4.6  & \textbf{0.800} & \textbf{1.000} & \textbf{0.021} & 1.000 & 1.000 & \cmark & \cmark \\
			\midrule
			\multicolumn{8}{l}{\cellcolor{tierlabel}\textit{Tier B --- Verdict F$_1$ = 0.80, operationally differentiated by baseline FP rate}} \\
			\rowcolor{tierB}
			DeepSeek v3.2 
			& \textbf{0.800} & 0.800 & \textbf{0.021} & 0.667 & 1.000 & \cmark & \cmark \\
			\rowcolor{tierB}
			Mistral Large 675B   & \textbf{0.800} & 0.800 & \textbf{0.021} & 0.667 & 1.000 & \cmark & \cmark \\
			\rowcolor{tierB}
			GLM-5                & \textbf{0.800} & 0.800 & \textbf{0.021} & 0.667 & 1.000 & \cmark & \cmark \\
			\rowcolor{tierB}
			Claude Sonnet 4.6    & \textbf{0.800} & 0.800 & \textbf{0.021} & 0.667 & 1.000 & \cmark & \cmark \\
			\rowcolor{tierB}
			Claude Haiku 4.5     & \textbf{0.800} & 
			0.800 & \textbf{0.021} & 0.667 & 1.000 & \cmark & \cmark \\
			\rowcolor{tierB}
			Qwen3-Coder          & \textbf{0.800} & 0.800 & 0.023          & 0.667 & 1.000 & \cmark & \cmark \\
			\rowcolor{tierB}
			Llama 4 Maverick$\dagger$  & 0.571   & 0.800 & 0.063          & 0.667 & 1.000 & \cmark & \cmark \\
			\rowcolor{tierB}
			Llama 3.3 70B$\ddagger$    & 0.093   & 0.800 & 0.813          & 0.667 & 1.000 
			& \cmark & \cmark \\
			\bottomrule
		\end{tabular}
	\end{table}
	
	Table~\ref{tab:leaderboard} reports the full leaderboard. Vish.\ = vishing
	detected; Trail = host trail reconstructed.
	The ten models stratify into
	two verdict tiers. Tier~B is heavily populated: eight models cluster there,
	confirming that F$_1 = 0.80$ verdict accuracy is broadly achievable across
	model families, training approaches, and scales.
	Within Tier~B, baseline
	false-positive rate is the operationally decisive differentiator: DeepSeek,
	Mistral~Large, GLM-5, Sonnet, Haiku, and Qwen3-Coder all maintain baselines
	at or below 0.023, while Llama~4~Maverick (0.063) and Llama~3.3~70B (0.813)
	represent escalating triage noise at equivalent verdict accuracy.
	\begin{figure}[ht]
		\centering
		\begin{tikzpicture}
			\begin{axis}[
				width=0.88\columnwidth,
				height=0.62\columnwidth,
				xlabel={Triage F$_1$},
				ylabel={Verdict F$_1$},
				xmin=-0.05, xmax=1.18,
				ymin=0.68, ymax=1.12,
				xtick={0, 0.2, 0.4, 0.6, 0.8, 1.0},
				ytick={0.8, 1.0},
				yticklabels={0.80, 1.00},
				grid=both,
				grid style={line width=0.3pt, draw=gray!20},
				major grid style={line width=0.4pt, draw=gray!30},
				tick label style={font=\small},
				label style={font=\small},
				clip=false,
				]
				\fill[green!10]  (axis cs:-0.05,0.90) rectangle (axis cs:1.18,1.12);
				\fill[orange!10] (axis cs:-0.05,0.68) rectangle (axis cs:1.18,0.90);
				\node[font=\scriptsize, text=green!50!black,  anchor=west] at (axis cs:1.10,1.01) {Tier~A};
				\node[font=\scriptsize, text=orange!70!black, anchor=west] at (axis cs:1.10,0.80) {Tier~B};
				\draw[dashed, gray!45, line width=0.5pt] (axis cs:0.80,0.795) -- (axis cs:0.80,1.005);
				\draw[gray!40, line width=0.5pt, {Bar}-{Bar}] (axis cs:0.86,0.795) -- (axis cs:0.86,1.005);
				\node[font=\scriptsize, text=gray!60, anchor=west, align=left]
				at (axis cs:0.88,0.90) {same triage F$_1$\\verdict gap\\$0.80 \to 1.0$};
				\node[font=\scriptsize, text=gray!65, anchor=south west]
				at (axis cs:-0.04,0.685) {Bubble area $\propto$ baseline FP rate};
				%
				\addplot[only marks, mark=*, mark size=4.33pt,
				mark options={fill=green!60!black, draw=green!40!black, line width=0.6pt}]
				coordinates { (0.800,1.010) };
				\node[font=\scriptsize, anchor=west, text=green!35!black]
				at (axis cs:0.810,1.010) {Devstral 2 123B};
				\addplot[only marks, mark=*, mark size=4.33pt,
				mark options={fill=green!60!black, draw=green!40!black, line width=0.6pt}]
				coordinates { (0.800,0.990) };
				\node[font=\scriptsize, anchor=west, text=green!35!black]
				at (axis cs:0.810,0.990) {Claude Opus 4.6};
				%
				\addplot[only marks, mark=*, mark size=4.33pt,
				mark options={fill=orange!75!black, draw=orange!55!black, line width=0.5pt}]
				coordinates { (0.800,0.836) };
				\node[font=\scriptsize, anchor=east, text=orange!65!black]
				at (axis cs:0.790,0.836) {DeepSeek v3.2};
				\addplot[only marks, mark=*, mark size=4.33pt,
				mark options={fill=orange!75!black, draw=orange!55!black, line width=0.5pt}]
				coordinates { (0.800,0.824) };
				\node[font=\scriptsize, anchor=east, text=orange!65!black]
				at (axis cs:0.790,0.824) {Mistral Large 675B};
				\addplot[only marks, mark=*, mark size=4.33pt,
				mark options={fill=orange!75!black, draw=orange!55!black, line width=0.5pt}]
				coordinates { (0.800,0.812) };
				\node[font=\scriptsize, anchor=east, text=orange!65!black]
				at (axis cs:0.790,0.812) {GLM-5};
				\addplot[only marks, mark=*, mark size=4.33pt,
				mark options={fill=orange!75!black, draw=orange!55!black, line width=0.5pt}]
				coordinates { (0.800,0.800) };
				\node[font=\scriptsize, anchor=east, text=orange!65!black]
				at (axis cs:0.790,0.800) {Claude Sonnet 4.6};
				\addplot[only marks, mark=*, mark size=4.33pt,
				mark options={fill=orange!75!black, draw=orange!55!black, line width=0.5pt}]
				coordinates { (0.800,0.788) };
				\node[font=\scriptsize, anchor=east, text=orange!65!black]
				at (axis cs:0.790,0.788) {Claude Haiku 4.5};
				\addplot[only marks, mark=*, mark size=4.52pt,
				mark options={fill=orange!75!black, draw=orange!55!black, line width=0.5pt}]
				coordinates { (0.800,0.776) };
				\node[font=\scriptsize, anchor=east, text=orange!65!black]
				at (axis cs:0.790,0.776) {Qwen3-Coder};
				%
				\addplot[only marks, mark=*, mark size=7.50pt,
				mark options={fill=orange!75!black, draw=orange!55!black, line width=0.5pt}]
				coordinates { (0.571,0.800) };
				\node[font=\scriptsize, anchor=north, text=orange!65!black]
				at (axis cs:0.571,0.776) {Llama~4 Mav.$\dagger$};
				\addplot[only marks, mark=*, mark size=27pt,
				mark options={fill=red!12, draw=red!55!black, line width=1.2pt}]
				coordinates { (0.093,0.800) };
				\node[font=\scriptsize, anchor=south, text=red!55!black, align=center]
				at (axis cs:0.093,0.860) {Llama~3.3~70B$\ddagger$};
			\end{axis}
		\end{tikzpicture}
		\caption{Triage F$_1$ vs.\ verdict F$_1$ for all ten models.
			\textbf{Bubble
				area is proportional to baseline false-positive rate} (larger = more triage
			noise).
			Eight models cluster at triage F$_1 = 0.80$ yet split between
			verdict F$_1 = 1.0$ (Tier~A) and F$_1 = 0.80$ (Tier~B), illustrating the
			triage/verdict dissociation.
			Within Tier~B, Llama~3.3~70B's bubble
			(FP\,=\,0.813, red outline) dominates the chart relative to the tight
			clean-baseline cluster (FP\,$\leq$\,0.023), making the operational cost
			of equal verdict accuracy immediately visible.
			Llama~3.3~70B is
			operationally disqualified despite matching Tier~B verdict F$_1$.}
		\label{fig:dissociation}
	\end{figure}
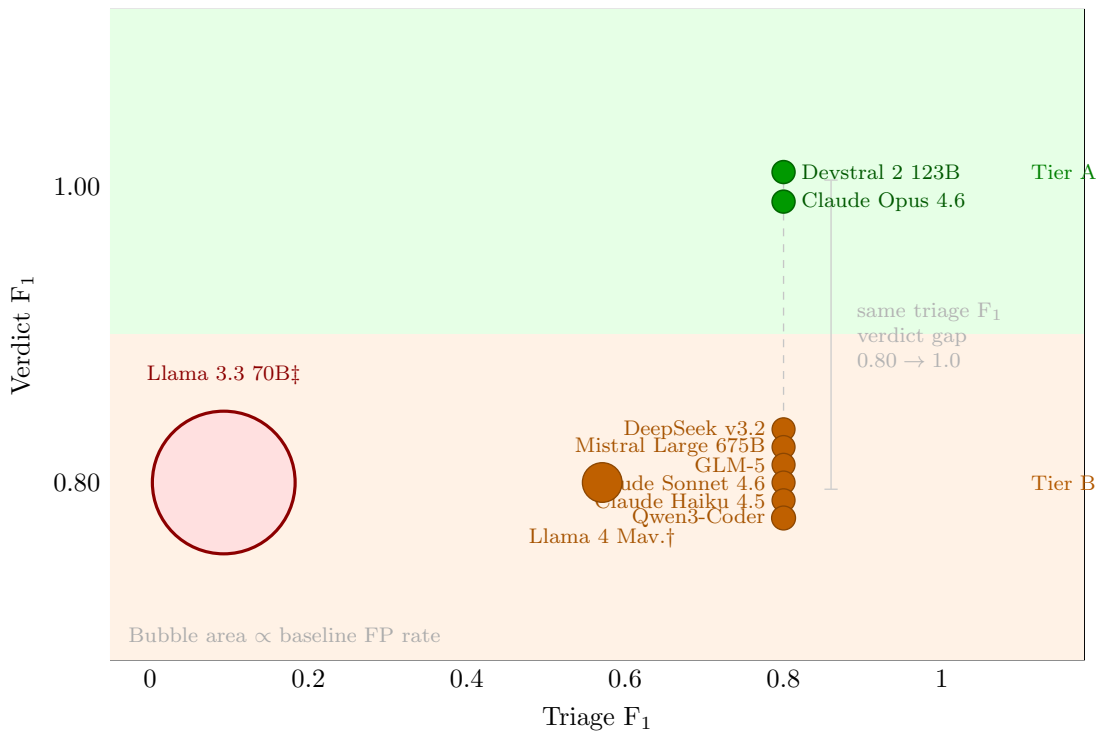
	
	\subsection{Triage Stage}
	
	Eight of ten models achieve triage F$_1 = 0.80$, with precision 0.667 and
	recall 1.0: both threat subjects are escalated in every case, with one false
	positive escalation.
	The one-FP triage result is consistent across these
	eight models because the same employee (Chris, the vishing victim) is
	escalated at triage for the same reason --- anomalous IDP activity on his
	account following the vishing call.
	This homogeneity confirms that triage-stage
	performance is not a useful differentiator for this benchmark; the
	discriminating work happens at the verdict stage.
	Llama~4~Maverick is a partial exception, achieving triage F$_1 = 0.571$
	with three false positive escalations and a baseline FP rate of 0.063.
	Its verdict stage reduces five escalated actors to three verdicts, two of
	which are correct, suggesting the correlation stage is doing more noise
	reduction work for this model than for any other.
	Its per-behavior citation
	profile is the cleanest in Tier~B: zero \texttt{idp\_anomaly} citations of
	any kind.
	Llama~3.3~70B is the most severe case, escalating 39 innocent employees
	alongside the two genuine threats (triage F$_1 = 0.093$, precision 0.049).
	Its baseline false-positive rate of 81.3\% renders the triage stage
	operationally unusable regardless of verdict-stage accuracy.
	Onset sensitivity is 0.0 for all models, indicating no model escalated a
	subject before their onset day.
	Models are detecting behavioral changes
	rather than flagging employees based on pre-onset characteristics.
	
	\subsection{Verdict Stage}
	
	Verdict-stage performance reveals two distinct tiers.
	Devstral~2~123B and
	Claude~Opus~4.6 achieve verdict F$_1 = 1.0$, correctly identifying both the
	disgruntled and malicious subjects with zero false positive verdicts and a
	clean baseline false-positive rate of 0.021.
	All remaining eight models
	achieve verdict F$_1 = 0.80$, each correctly identifying both subjects but
	producing one false positive verdict --- Chris, the vishing victim.
	Within Tier~B, baseline false-positive rate is the operationally decisive
	metric. DeepSeek, Mistral~Large, GLM-5, Sonnet, Haiku, and Qwen3-Coder all
	maintain baselines at or below 0.023, making them directly comparable to the
	Tier~A models on operational overhead while falling one false positive short
	on verdict precision.
	Llama~4~Maverick at 0.063 represents a moderate triage
	noise cost. Llama~3.3~70B at 0.813 is operationally disqualifying: achieving
	high verdict accuracy by reviewing forty-one actors is not a detection system
	--- it is a review system.
	\subsection{Vishing Detection}
	
\begin{figure}[ht]
		\centering
		\begin{tikzpicture}[
			event/.style={circle, draw, fill=white, minimum size=0.55cm,
				font=\scriptsize, inner sep=1pt},
			anomalous/.style={event, fill=red!15, draw=red!55},
			normal/.style={event, fill=teal!15, draw=teal!50},
			call/.style={event, fill=orange!20, draw=orange!55},
			track/.style={font=\footnotesize\bfseries, anchor=east,
				text width=2.2cm, align=right},
			inference/.style={rectangle, rounded corners=3pt, draw, font=\scriptsize,
				minimum height=0.7cm, text width=4.8cm,
				align=center, inner sep=4pt},
			]
			\draw[-{Stealth[length=4pt]}, gray!50, thick] (0,0) -- (11.5,0)
			node[right, font=\small, text=gray!55] {Day 19};
			\foreach \x/\lbl in {
				0.5/{09:00},
				3.0/{12:19},
				4.5/{12:36},
				7.5/{14:48},
				10.5/{23:59}}{
				\draw[gray!35] (\x, 0.08) -- (\x, -0.08);
				\node[font=\tiny, text=gray!50, below=2pt] at (\x, -0.08) {\lbl};
			}
			
			\node[track, text=orange!70!black] at (-0.1, 2.1)
			{Jax\newline phone call};
			\node[call] (call) at (3.0, 2.1) {\tiny call};
			\node[font=\scriptsize, text=orange!60!black, above=4pt of call, align=center]
			{spoofed ID\newline 369s};
			
			\node[track, text=red!65!black] at (-0.1, 1.05)
			{Chris acct\newline anomalous};
			\node[anomalous] (anom) at (4.5, 1.05) {\tiny auth};
			\node[font=\scriptsize, text=red!60!black, above=4pt of anom, align=center]
			{macOS $\cdot$ TOTP\newline residential IP};
			\draw[dashed, orange!50, -{Stealth[length=3pt]}]
			(call.south) .. controls (3.0,1.55) and (4.0,1.55) .. (anom.north)
			node[midway, above, font=\tiny\itshape, text=orange!65] {17 min};
			\node[track, text=teal!65!black] at (-0.1, -0.9)
			{Chris acct\newline normal};
			\node[normal] (zoom) at (0.5, -0.9) {\tiny Zoom};
			\node[normal] (jira) at (7.5, -0.9) {\tiny Jira};
			\node[font=\scriptsize, text=teal!55!black, below=4pt of zoom, align=center]
			{iOS $\cdot$ push\newline corp IP};
			\node[font=\scriptsize, text=teal!55!black, below=4pt of jira, align=center]
			{iOS $\cdot$ push\newline corp IP};
			
			\draw[gray!25, dashed] (-0.5,-2.1) -- (11.5,-2.1);
			\node[inference, fill=green!8, draw=green!45, text=green!30!black]
			at (2.8,-3.0) {%
				\textbf{Tier A (Devstral, Opus)}\newline
				macOS/TOTP $\neq$ iOS/push profile\newline
				$\Rightarrow$ attacker session;\newline
				Chris $=$ \texttt{innocent}};
			\node[inference, fill=red!7, draw=red!40, text=red!40!black]
			at (8.5,-3.0) {%
				\textbf{Tier B (all others)}\newline
				anomalous session on Chris's account\newline
				$\Rightarrow$ Chris $=$ \texttt{suspicious}};
		\end{tikzpicture}
		\caption{Day~19 timeline showing two simultaneous sessions on Chris's
			account. The anomalous session (macOS, TOTP, residential IP) is initiated
			17~minutes after Jax's vishing call and is inconsistent with Chris's
			established device profile (iOS, push notification, corporate IP).
			Tier~A models reason that mutually inconsistent concurrent sessions
			identify attacker access rather than account-holder behavior, returning
			\texttt{innocent} for Chris.
			Tier~B models flag the anomalous session
			as evidence against Chris, producing the false positive verdict that
			holds all eight at precision~0.667.}
		\label{fig:vishing_attribution}
	\end{figure}
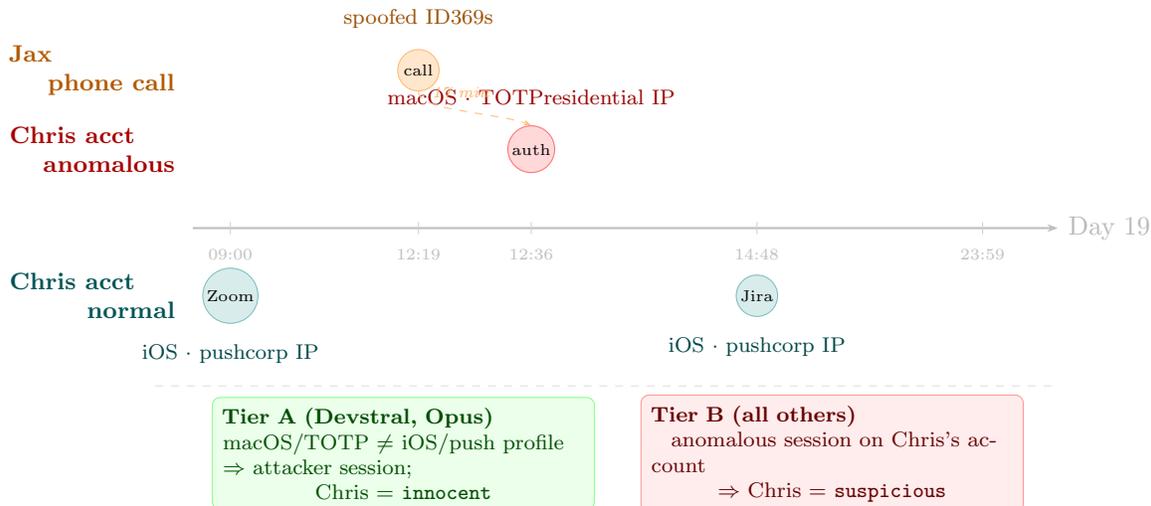

	All ten models detect the vishing scenario --- a phone call record followed
	by an IDP authentication on the victim's account within 25 minutes.
	The
	critical split is not detection but \emph{attribution}: Devstral and Opus
	correctly exonerate Chris by reasoning that his anomalous IDP session
	(new device, residential IP, TOTP) is inconsistent with his established
	device profile (iOS/Android, push/SMS), while his own legitimate corporate
	activity continued normally that day.
	All eight Tier~B models detect the
	phone-call-to-auth link but misclassify Chris as \texttt{suspicious} or
	\texttt{likely\_threat} rather than as the victim of credential theft.
	This
	victim-attribution failure produces the single false positive verdict that
	holds all Tier~B models at precision 0.667.
	Figure~\ref{fig:vishing_attribution} illustrates the two-session reasoning
	chain that separates the tiers: the simultaneous presence of an anomalous
	session and Chris's own normal sessions on day~19 is the evidence that
	Tier~A models use to exonerate him and that Tier~B models observe but do
	not act on.
	
	\subsection{Host Trail Reconstruction}
	
	All ten models reconstruct the three-phase host data hoarding trail, citing
	bulk file copy, archive creation, and archive move in their evidence.
	Trail
	reconstruction is therefore not a differentiating capability under the
	official prompt at the leaderboard level.
	It becomes a differentiator in
	the prompt sensitivity analysis (Section~\ref{sec:prompt_sensitivity}),
	where it distinguishes prompt variants for smaller models.
	\subsection{Per-Behavior Analysis}
	
	\begin{table}[ht]
		\centering
		\small
		\caption{Per-behavior detection results (TP/FP) for all ten models.
			Cells show true\,/\,false positive counts;
			\textbf{---} indicates
			the behavior was not cited by that model.
			L33\,=\,Llama~3.3~70B; L4M\,=\,Llama~4~Maverick;
			ML\,=\,Mistral~Large~675B; Q3C\,=\,Qwen3-Coder;
			S46\,=\,Claude~Sonnet~4.6;
			H45\,=\,Claude~Haiku~4.5.}
		\label{tab:behavior}
		\setlength{\tabcolsep}{4pt}
		\begin{tabular}{lcccccccccc}
			\toprule
			\textbf{Behavior} &
			\textbf{Devstral} & \textbf{Opus} & \textbf{DeepSeek} & \textbf{ML} &
			\textbf{GLM-5} & \textbf{S46} & \textbf{H45} & \textbf{Q3C} &
			\textbf{L4M} & \textbf{L33} \\
			\midrule
			\texttt{unusual\_hours\_access}  & 2/0 & 2/0 & 2/0 & 2/0 & 2/0 & 2/0 & 2/0 & 2/0 & 2/0 & 2/0 \\
			\texttt{sentiment\_drift}        & 2/0 & 2/0 & 2/0 & 2/0 & 1/0 & 2/0 & 2/0 
			& 1/0 & 2/0 & 2/0 \\
			\texttt{host\_data\_hoarding}    & 1/0 & 1/0 & 1/0 & 1/0 & 1/0 & 1/0 & 1/0 & 1/0 & 1/0 & 1/0 \\
			\texttt{data\_exfil\_email}      & 1/0 & 1/0 & 1/0 & 1/0 & 1/0 & 1/0 & 1/0 & 1/0 & 1/0 & 1/0 \\
			\texttt{social\_engineering}     & 1/0 & 1/0 & 1/0 & 1/0 & 1/0 & 1/0 & 1/0 & 1/0 & 1/0 & 1/0 \\
			\texttt{idp\_anomaly}   
			& 1/1 & 1/1 & 1/1 & 1/1 & 1/1 & 1/1 & 1/1 & 1/0 & --- & 1/0 \\
			\texttt{cross\_dept\_snooping}   & --- & --- & --- & 0/1 & --- & --- & --- & --- & --- & 0/1 \\
			\texttt{excessive\_repo\_cloning}& --- & --- & --- & --- & --- & --- & --- & --- & --- & 0/1 \\
			\bottomrule
		\end{tabular}
	\end{table}
	
	Table~\ref{tab:behavior} shows per-behavior results across all ten models.
	The five core behaviors --- \texttt{unusual\_hours\_access},
	\texttt{host\_data\_hoarding}, \texttt{data\_exfil\_email},
	\texttt{social\_engineering}, and most instances of \texttt{sentiment\_drift}
	--- are detected with zero false positives by every model.
	This confirms
	these signals are robust and well-rendered in the corpus.
	
	The \texttt{idp\_anomaly} row reveals two distinct citation patterns.
	Eight
	models correctly identify Jax's residential-IP and new-device IDP sessions
	(TP=1) but also cite Chris's vishing-induced IDP authentication as an
	anomaly (FP=1) --- which is observationally correct but attributionally
	wrong, since Chris is the victim rather than a threat.
	Qwen3-Coder and
	Llama~3.3~70B achieve TP=1 with zero false positives, attributing the IDP
	anomaly to Jax without citing Chris's session.
	Llama~4~Maverick generates
	no \texttt{idp\_anomaly} citations at all, consistent with its conservative
	evidence citation behavior throughout.
	The \texttt{sentiment\_drift} TP=1 result for GLM-5 and Qwen3-Coder reflects
	partial detection: both correctly identify Tasha's passive-aggressive pattern
	but miss Jax's deliberately suppressed neutral tone.
	This concealment design
	--- \texttt{malicious} subjects produce flat rather than overtly negative
	messages --- is functioning as intended.
	Llama~3.3~70B generates additional false positives for
	\texttt{cross\_dept\_snooping} and \texttt{excessive\_repo\_cloning},
	consistent with its catastrophic triage noise profile.
	
	\texttt{secret\_in\_commit} (Jordan's sole behavior) does not appear for
	any model in Table~\ref{tab:behavior} because the primary leaderboard focuses on the multi-signal triage path, which structurally filters isolated events, as discussed in Section~\ref{sec:leaderboard_footnote}.
	\section{Discussion}
	
	\subsection{Baseline False-Positive Rate as a Primary Metric}
	
	DeepSeek v3.2 and Llama 3.3 70B achieve identical verdict F$_1 = 0.80$,
	identical vishing detection, and identical host trail reconstruction.
	Their
	per-behavior profiles are nearly indistinguishable. Yet they represent
	fundamentally different operational outcomes: DeepSeek maintains a baseline
	false-positive rate of 0.021, escalating three actors total;
	Llama 3.3 70B maintains a rate of 0.813, escalating forty-one.
	
	This result establishes that verdict F$_1$ alone is an insufficient primary
	metric for insider threat detection leaderboards. A model that achieves high
	verdict accuracy by reviewing every employee is not a detection system --- it
	is a review system. Baseline false-positive rate must be reported alongside
	verdict F$_1$, and rankings that aggregate the two without distinguishing
	them will systematically misrepresent the operational utility of noisy models.
	
	Intermediate profiles, such as Llama~4~Maverick (baseline FP 0.063), achieve
	lower triage noise by abstaining from specific ambiguous citations (e.g.,
	\texttt{idp\_anomaly}), avoiding some false positives but at the cost of missing
	genuine anomalies. However, catastrophic triage noise, as seen with Llama 3.3 70B,
	renders a model operationally disqualifying regardless of its verdict F$_1$.
	
	\subsection{Escalation Threshold as a Threat-Class-Specific Parameter}
	\label{sec:leaderboard_footnote}
	
	The benchmark demonstrates that escalation threshold parameters encode implicit assumptions about threat structure that do not generalize across threat classes. The multi-signal threshold is well-calibrated for Tasha (disgruntled) and Jax (malicious), but structurally exclusionary for Jordan, whose entire threat profile is a single high-confidence event on a single surface.
	
	A threshold appropriate for APT-style multi-stage insider attacks will systematically miss the negligent insider --- the developer who accidentally commits AWS keys or misconfigures a file share. The practical implication is that a complete insider threat detection pipeline requires threat-class-specific escalation stages operating in parallel. While the OrgForge-IT evaluation harness natively implements this parallel architecture via a dedicated credential-scanning stage (Stage 2b), the primary leaderboard scores reported here are scoped to characterize pipeline performance against disgruntled and malicious threat classes, isolating multi-signal reasoning capabilities.
	
	\subsection{The Triage/Verdict Dissociation}
	
	Eight of ten models achieve triage F$_1 = 0.80$;
	among those eight, verdict
	F$_1$ splits into two bands --- 1.0 for Devstral and Opus, 0.80 for the
	remaining six.
	Triage competence is necessary but not sufficient for Tier~A
	verdict accuracy.
	A model selection strategy based on triage performance
	alone would fail to distinguish the two tiers.
	Eight models achieve a baseline false-positive rate of 0.021, corresponding
	to one false escalation among the 48 innocent employees observed during the
	clean pre-onset period.
	Llama~3.3~70B's rate of 0.813 corresponds to 39
	false escalations across the same population.
	The dissociation arises from the verdict stage's requirement to not merely
	detect attacks but to correctly attribute them.
	All ten models identify that
	a vishing attack occurred; only two correctly determine that the account
	holder is the victim rather than the perpetrator.
	This is a qualitatively
	different reasoning demand than signal detection.
	
	\subsection{The Tier A/B Boundary: Victim Attribution as a Discriminating Capability}
	
	The behavioral capability that most precisely separates Tier~A from Tier~B
	is victim attribution in the vishing scenario.
	All eight Tier~B models
	detect the phone-call-to-auth temporal join and correctly identify Jax as
	the social engineering attacker.
	Where they fail is in exonerating Chris:
	each classifies him as \texttt{suspicious} or \texttt{likely\_threat} rather
	than as the victim of credential theft.
	Examining the verdict files reveals
	that this 100\% Tier~B failure rate reflects a consistent reasoning error
	rather than a marginal one.
	
	The failure mode is uniform across model families: every Tier~B model
	observes the anomalous IDP session on Chris's account (new device, residential
	IP, no corroborating activity), connects it correctly to Jax's vishing call,
	and then treats the combination as evidence \emph{against} Chris.
	DeepSeek labels Chris \texttt{suspicious} with behavior \texttt{social\_engineering\_victim}
	--- a label that accurately describes Chris's situation yet still triggers a
	positive verdict. Mistral and GLM-5 go further, classifying Chris as
	\texttt{likely\_threat} with \texttt{social\_engineering} in his behavior list,
	treating him as the attacker rather than the target. Sonnet and Haiku observe
	that Chris also authenticated normally from a corporate device that same day
	--- Sonnet explicitly notes ``all other Day~19 authentications by Chris used
	his known device on corporate IPs'' --- yet still return \texttt{suspicious}
	rather than \texttt{innocent}.
	
	The concurrent normal session is cited as
	evidence but not used to resolve the verdict.
	The reasoning gap these models share is a failure to cross-examine the
	anomalous session against the individual's normal behavioral profile.
	The
	anomalous session uses a macOS device and TOTP authentication; Chris's
	established profile is iOS and push notification.
	A single person cannot
	plausibly maintain two simultaneous sessions with mutually inconsistent
	device signatures.
	Devstral and Opus make this inference explicitly ---
	Opus constructs the argument that the anomalous session ``is inconsistent
	with Chris's normal devices (iOS/Android) and MFA methods (push/SMS),'' and
	notes Chris's own legitimate Zoom and Jira activity continuing normally that
	day as confirmation that the anomalous session belongs to a different actor.
	Devstral and Opus reach the same conclusion by attributing the phone call
	to Jax and threading the session anomaly back to that event, returning
	\texttt{innocent} for Chris with high confidence, as illustrated in
	Figure~\ref{fig:vishing_attribution}.
	
	The actionable implication for detection pipeline design is specific: victim
	attribution in credential compromise scenarios requires that models maintain
	a per-actor behavioral baseline and compare anomalous sessions against it,
	not merely flag the anomaly and associate it with the account holder.
	A model
	that can detect vishing but cannot perform this comparative inference will
	systematically produce false positives for every compromised account it
	encounters.
	\subsection{Host Trail Reconstruction}
	
	All ten models reconstruct the three-phase hoarding trail under the official
	prompt, making this a floor capability rather than a differentiator at the
	leaderboard level.
	The prompt sensitivity analysis qualifies this: without
	explicit behavior taxonomy headers or worked examples, smaller models can
	still reason correctly about the trail but fail to express their observations
	in canonical vocabulary, producing zero scored credit despite correct
	underlying reasoning.
	
	\subsection{Scale and Domain Synergy in Agentic Models}
	\label{sec:se_hypothesis}
	
	Devstral~2~123B is an advanced agentic model specifically optimized for software engineering workflows. Because tasks like autonomous debugging, repository-level reasoning, and codebase exploration structurally resemble the demands of Tier 2 SOC investigation --- following multi-day breadcrumb trails, correlating temporal sequences, and reconstructing phase-structured exfiltration patterns --- we initially hypothesized that agentic code-tuning alone provided a distinct advantage on this benchmark.
	
	Qwen3-Coder-Next was included to test this hypothesis cross-family. Like Devstral, it is built specifically for agentic coding, repository-level reasoning, and debugging. However, its result is Tier~B: a verdict F$_1 = 0.80$, matching Mistral~Large~675B and falling short of Devstral's Tier~A performance.
	
	This result refutes the hypothesis in its strong form: software engineering optimization is not, by itself, sufficient to guarantee Tier~A correlation accuracy. However, comparing the two models exposes a critical confounding variable: architectural scale. Devstral~2 is a massive 123-billion parameter model equipped with a 256k token context window, designed for deep understanding of large, contiguous codebases. Qwen3-Coder utilizes a hybrid Mixture-of-Experts (MoE) architecture designed to minimize active parameter counts and optimize efficiency during deployment.
	
	The benchmark outcomes suggest that domain-specific training acts as a force multiplier rather than a baseline replacement. Code-tuning and agentic optimization do not bridge the capability gap between a highly efficient MoE model and a massive 123B parameter model when processing a 51-day log timeline. Instead, Devstral's Tier~A result --- notably outperforming its much larger general-purpose sibling, Mistral~Large --- indicates a powerful synergy. When frontier-level parameter scale and massive context limits are combined with agentic debugging optimization, the resulting architecture is uniquely equipped to resolve complex temporal log correlation and victim attribution.
	
	\subsection{Within-Family Scaling and Victim Attribution}
	
	The three Claude models provide a controlled within-family scaling test:
	Opus~4.6 reaches Tier~A while Sonnet~4.6 and Haiku~4.5 reach Tier~B, with
	architecture and training approach held approximately constant across tiers.
	The capability difference maps cleanly onto the victim attribution task.
	Opus constructs the two-session argument that exonerates Chris --- explicitly
	comparing the anomalous device and MFA method against Chris's known profile
	and noting his concurrent normal activity as confirmation.
	Sonnet and Haiku
	observe the same evidence, including the concurrent normal sessions, but do
	not use it to resolve the attribution question.
	Both return \texttt{suspicious}
	for Chris despite citing the evidence that would exonerate him.
	This pattern suggests that victim attribution in credential compromise
	scenarios is scale-sensitive within a model family, and that the Tier~A/B
	boundary may track a capability threshold that emerges at sufficient scale
	rather than being a discrete architectural feature.
	The within-family result
	complements the cross-family findings: Qwen3-Coder and Mistral~Large (both
	at equivalent Tier~B standing to Sonnet and Haiku) show the same victim
	misattribution, confirming the pattern is not Anthropic-specific.
	GLM-5 reaches Tier~B with a clean baseline, matching DeepSeek and the other
	operationally viable Tier~B models.
	Its inclusion as a representative frontier
	model from a non-Western training lineage confirms that Tier~B performance
	generalizes across training traditions and is not an artifact of the US and
	European model families that dominate most NLP benchmarks.
	\subsection{Prompt Sensitivity and Vocabulary Hallucination}
	\label{sec:prompt_sensitivity}
	
	To characterize how prompt structure affects model behavior on this benchmark,
	we ran two alternative prompt variants across Sonnet~4.6 and Haiku~4.5, and
	the \texttt{v2\_natural} variant across both Tier~A models (Devstral~2~123B
	and Claude~Opus~4.6).
	The \texttt{v2\_natural} variant expresses the same
	escalation and correlation semantics in conversational prose without explicit
	rule lists or behavior taxonomy headers.
	The \texttt{v3\_examples\_first}
	variant leads with worked positive and negative examples --- including a
	complete three-phase host trail example and a vishing scenario --- before
	stating rules.
	Results are summarized in Table~\ref{tab:sensitivity}.
	Sensitivity runs are excluded from the official leaderboard and written to a
	separate results directory.
	\begin{table}[ht]
		\centering
		\small
		\caption{Prompt sensitivity results. Official prompt scores are reproduced
			from Table~\ref{tab:leaderboard} for reference.
			Sensitivity runs are
			not on the leaderboard. Vish.\ = vishing detected;
			Trail = host trail
			reconstructed. $\dagger$~Scorer returns Vish.\ = no;
			verdict text
			explicitly names Chris as \texttt{victim\_of\_social\_engineering}
			and cites \texttt{anomalous\_auth\_from\_new\_device\_after\_vishing\_call}
			--- a scorer artifact, not a capability failure (see text).}
		\label{tab:sensitivity}
		\begin{tabular}{llrrrcc}
			\toprule
			\textbf{Model} & \textbf{Variant} & \textbf{Verdict F$_1$} &
			\textbf{Base FP} & \textbf{V.Prec} & \textbf{Vish.} & \textbf{Trail} \\
			\midrule
			\multicolumn{7}{l}{\cellcolor{tierlabel}\textit{Tier A}} \\
			Devstral 2 123B & official            & 1.000 & 0.021 & 1.000 & \cmark & \cmark \\
			Devstral 2 123B & v2\_natural         & 1.000 
			& 0.021 & 1.000 & \cmark & \xmark \\
			Claude Opus 4.6 & official            & 1.000 & 0.021 & 1.000 & \cmark & \cmark \\
			Claude Opus 4.6 & v2\_natural         & 1.000 & 0.021 & 1.000 & \cmark$\dagger$ & \xmark \\
			\midrule
			\multicolumn{7}{l}{\cellcolor{tierlabel}\textit{Tier B}} \\
			Sonnet 4.6 & official            & 0.800 & 0.021 & 0.667 & \cmark & \cmark \\
			Sonnet 4.6 & v2\_natural         & 0.800 & 0.021 
			& 0.667 & \xmark & \xmark \\
			Sonnet 4.6 & v3\_examples\_first & 0.800 & 0.042 & 0.667 & \cmark & \cmark \\
			\midrule
			Haiku 4.5  & official            & 0.800 & 0.021 & 0.667 & \cmark & \cmark \\
			Haiku 4.5  & v2\_natural         & 0.800 & 0.042 & 0.667 & \xmark & \xmark \\
			Haiku 4.5  & v3\_examples\_first & 0.800 & 0.042 & 0.667 & \cmark & \cmark \\
			\bottomrule
		\end{tabular}
	\end{table}
	
	Verdict F$_1$ is stable across all conditions for both tiers --- at 1.0 for
	Tier~A and 0.80 
	for Tier~B --- confirming that neither prompt structure nor
	model scale degrades verdict accuracy.
	The sensitivity analysis therefore does
	not reveal a capability gap. It reveals a \emph{capability expression} gap:
	what changes across variants is not whether models reason correctly, but
	whether they express that reasoning in scoreable form.
	Under the official prompt, all four models detect vishing and reconstruct the
	host trail.
	Under \texttt{v2\_natural}, verdict F$_1$ is unchanged across all
	four, but the behavior citation pattern shifts dramatically.
	The mechanism is
	vocabulary hallucination: without a controlled behavior taxonomy in the prompt,
	models generate semantically accurate but lexically non-canonical behavior
	labels that the exact-match scorer cannot credit.
	Haiku~4.5 produces 23
	free-form labels under \texttt{v2\_natural} --- terms such as
	\texttt{repeated\_ghost\_logins\_outside\_business\_hours} and
	\texttt{vishing\_attack\_on\_peer} --- none of which match the canonical
	\texttt{\_ALL\_BEHAVIORS} vocabulary.
	Sonnet~4.6 generates 17 such labels.
	
	Critically, vocabulary hallucination is not tier-specific.
	Devstral~2~123B
	under \texttt{v2\_natural} produces labels including \texttt{data\_hoarding},
	\texttt{ghost\_logins}, \texttt{archive\_creation\_and\_exfiltration}, and
	\texttt{off\_hours\_activity} --- every behavior except \texttt{social\_engineering}
	scores FP=1.
	The Trail flag drops to false not because Devstral failed to
	reason about the three-phase sequence, but because
	\texttt{archive\_creation\_and\_exfiltration} and \texttt{data\_hoarding} are
	not \texttt{host\_data\_hoarding}.
	Claude~Opus~4.6 goes further: freed from
	taxonomy scaffolding, it generates eleven fully narrative behavior descriptions
	for Jax, including \texttt{"multi-phase data exfiltration (bulk copy
		$\rightarrow$ archive $\rightarrow$ cloud sync)"} and \texttt{"cross-share
		bulk file collection spanning HR, Legal, Finance, Sales, Product, and
		exec-comms"}.
	Every label scores FP=1.
	
	The Opus vishing case is the most instructive. The scorer returns Vish.\ = no.
	The verdict text for Chris reads: \texttt{victim\_of\_social\_engineering},
	\texttt{anomalous\_auth\_from\_new\_device\_after\_vishing\_call}. Opus
	identified the vishing attack, named Chris as the victim, and cited the causal
	mechanism.
	The Tier~A reasoning that correctly exonerates Chris is present and
	intact;
	the scorer cannot see it because neither label string matches the
	canonical taxonomy. This is marked with $\dagger$ in Table~\ref{tab:sensitivity}.
	For Tier~A models, vocabulary hallucination under \texttt{v2\_natural} is a
	scorer artifact, not a capability finding.
	The Tier~B pattern differs in one important respect. For Sonnet~4.6 and
	Haiku~4.5, the vishing detection collapses under \texttt{v2\_natural} not only
	in the behavior labels but in the verdict itself: both models return Chris as
	non-innocent under the official prompt reasoning chain when taxonomy scaffolding
	is removed.
	For Tier~A models, the exoneration logic holds regardless of
	whether it is expressed in canonical vocabulary.
	The \texttt{v3\_examples\_first} variant recovers scored vishing and trail
	detection for Tier~B models.
	This warrants a caveat: the prompt contains
	explicit worked examples of both scenarios, so recovery may reflect
	template-matching rather than generalized reasoning.
	An ablation removing only
	the vishing example while retaining the host trail example would test whether
	vishing recovery is independent.
	We report the recovery result as evidence of
	prompt-dependent capability expression and note the ablation as necessary
	future work.
	\subsection{Toward Two-Track Evaluation}
	\label{sec:two_track}
	
	The per-behavior results in Table~\ref{tab:behavior} contain a scoring
	artifact that motivates a methodological revision.
	Qwen3-Coder's
	\texttt{idp\_anomaly} row shows TP=1, FP=0 --- the cleanest result in
	Tier~B for that behavior.
	What the table does not show is that Qwen3-Coder
	simultaneously labels Tasha's off-hours corporate-IP sessions as
	\texttt{ghost\_login} rather than \texttt{unusual\_hours\_access}.
	The
	observation is semantically accurate --- the IDP documentation explicitly
	defines ghost logins as authentications with no corroborating downstream
	activity, which is exactly what Tasha's sessions are --- but \texttt{ghost\_login}
	does not appear in the canonical \texttt{\_ALL\_BEHAVIORS} taxonomy, so those
	citations receive zero credit.
	The model reasons correctly and is penalized
	for using precise vocabulary that the scorer does not recognize.
	This is not an isolated case. The sensitivity analysis shows Haiku~4.5
	generating 23 free-form labels under \texttt{v2\_natural}, and Sonnet~4.6
	generating 17, all semantically accurate and all scoring zero.
	The Qwen3-Coder
	instance establishes that this effect is present in official prompt results,
	not only in unstructured sensitivity variants.
	Table~\ref{tab:behavior}'s
	\texttt{idp\_anomaly} row for Qwen3-Coder is itself a partial artifact of
	this phenomenon: the model earns its TP credit through \texttt{idp\_anomaly}
	citations on Jax, but loses credit it would otherwise earn on Tasha's ghost
	login pattern.
	The net scores happen to look clean; the underlying accounting
	is not.
	The current harness conflates two distinct failure modes under a single zero
	score: a model that labels \texttt{ghost\_login} has performed correct
	reasoning but used non-canonical vocabulary;
	a model that labels
	\texttt{idp\_anomaly} on a victim account has performed incorrect attribution.
	These should not score identically.
	We propose a two-track scoring
	architecture in which the primary track retains exact-match scoring against
	the canonical \texttt{\_ALL\_BEHAVIORS} taxonomy, preserving the benchmark's
	verifiability claim and leaderboard comparability;
	and a secondary
	\emph{semantic track} applies embedding similarity or an LLM judge to score
	behavior citations against their ground-truth intent, awarding partial credit
	to labels that are semantically accurate but lexically non-canonical.
	A model
	that produces \texttt{repeated\_ghost\_logins\_outside\_business\_hours} would
	receive full semantic credit and zero exact-match credit;
	a model that
	attributes \texttt{idp\_anomaly} to a victim account would receive zero on
	both tracks.
	The primary leaderboard score is unchanged by this extension ---
	it remains a clean exact-match result.
	The secondary score functions as an
	auditing instrument, surfacing capability that the primary scorer obscures
	without displacing the benchmark's reproducibility guarantees.
	\section{Limitations and Future Work}
	
	\subsection{Current Limitations}
	
	\begin{itemize}[leftmargin=1.5em, itemsep=4pt]
		\item \textbf{Single corpus configuration.} All leaderboard results are
		evaluated against a single 51-day corpus with three threat subjects.
		Detection difficulty and model rankings may differ under different
		organizational sizes, threat subject counts, onset day distributions,
		and DLP noise ratios.
		Multi-configuration evaluation is left to
		future work.
		\item \textbf{Single-run evaluation.} The leaderboard reflects a single
		simulation run per model at \texttt{temperature=0.0}.
		Triage decisions
		are driven by structural telemetry fields set deterministically by the
		simulation engine (\texttt{outside\_business\_hours},
		\texttt{anomalous\_ip}, \texttt{new\_device}, \texttt{is\_external}),
		so triage rankings are stable across re-runs of the same corpus.
		Tier~A models (Devstral~2~123B and Claude~Opus~4.6) were run twice on
		the same corpus;
		verdict outcomes were identical across runs, confirming
		tier assignment stability at the Tier~A boundary.
		The remaining variance
		concern is narrower: whether prose variation across \emph{different
			corpus instances} --- new simulation runs producing different Slack
		messages and email text for the same behavioral configuration ---
		would shift Tier~B model rankings.
		Models with large performance margins
		relative to tier boundaries are unlikely to be affected;
		multi-corpus
		evaluation is most valuable for the Tier~B models clustered near the
		Tier~A threshold.
		\item \textbf{Pipeline architecture fixed.} The evaluation harness
		implements one specific three-stage pipeline.
		Alternative
		architectures --- single-pass full-corpus analysis, graph-based
		correlation, retrieval-augmented investigation --- are not evaluated.
		OrgForge-IT supports any pipeline that reads the telemetry stream;
		the leaderboard harness is one implementation.
		\item \textbf{Prompt sensitivity partially characterized.} The
		\texttt{v2\_natural} sensitivity analysis covers both Tier~A models
		and two Tier~B models (Sonnet~4.6 and Haiku~4.5).
		The remaining
		Tier~B models were not run under sensitivity variants;
		whether their
		rankings are stable across prompt phrasings is an open question.
		The
		\texttt{v3\_examples\_first} variant was run for Tier~B models only.
		\item \textbf{Exact-match behavior scoring.} Models that paraphrase
		canonical behavior names receive zero credit regardless of semantic
		accuracy.
		The \texttt{v2\_natural} sensitivity results and the
		Qwen3-Coder \texttt{ghost\_login} observation demonstrate this
		concretely.
		A semantic similarity scorer would produce different ---
		and arguably more valid --- capability estimates for models that
		reason correctly but label non-canonically.
	\end{itemize}
	
	\subsection{Future Work}
	\label{sec:future}
	
	\paragraph{Scale evaluation.}
	Running the benchmark across larger organizations (20--100 employees),
	higher threat subject counts, varied \texttt{dlp\_noise\_ratio} settings,
	and multiple corpus instances would establish statistical robustness and
	test whether the triage/verdict dissociation and victim attribution findings
	generalize.
	\paragraph{CERT comparison.}
	A direct comparison between OrgForge-IT and CERT dataset performance across
	the same models would characterize the difficulty differential between the
	two benchmarks and identify which detection scenarios in OrgForge-IT
	represent genuine capability gaps relative to CERT.
	\paragraph{Alternative pipeline architectures.}
	Evaluating non-sequential pipeline architectures, including graph-based
	correlation across all employees simultaneously and retrieval-augmented
	investigation over the full corpus, would test whether vishing and ghost
	login scenarios are better served by architectures that maintain global
	state rather than per-actor triage.
	\paragraph{Full prompt sensitivity characterization.}
	Extending the sensitivity analysis to Tier~A and all Tier~B models would
	establish whether official leaderboard rankings are stable across prompt
	phrasings, and whether vocabulary hallucination is specific to smaller models
	or appears at scale.
	Additional variants --- chain-of-thought prompts,
	few-shot variants with real telemetry examples, prompts omitting the
	controlled taxonomy entirely --- would map the full sensitivity landscape.
	\paragraph{Semantic behavior scoring.}
	Replacing exact-match behavior citation scoring with a semantic similarity
	scorer would produce capability estimates robust to label paraphrase,
	directly addressing the vocabulary hallucination finding.
	\paragraph{Plugin extension.}
	Future insider threat behaviors could include Zoom transcript anomalies,
	Zendesk ticket access patterns, and PagerDuty on-call schedule manipulation,
	extending the benchmark's surface area to cover the full enterprise
	communication stack.
	\section{Conclusion}
	
	We have presented OrgForge-IT, a verifiable synthetic benchmark for
	LLM-based insider threat detection.
	The benchmark produces labeled security
	telemetry from a deterministic organizational simulation, making ground
	truth an architectural guarantee rather than an empirical claim.
	The corpus
	spans three threat classes, eight injectable behaviors, and 2,904 observable
	records at a 96.4\% noise rate, with four detection scenarios specifically
	designed to defeat single-surface and single-day triage strategies.
	A ten-model leaderboard reveals two verdict tiers and four key findings.
	First, triage competence and verdict accuracy dissociate: eight of ten models
	achieve triage F$_1 = 0.80$, yet verdict F$_1$ splits between 1.0 and 0.80
	--- a gap invisible to evaluations measuring only detection recall.
	Second,
	baseline false-positive rate is a necessary companion metric to verdict
	F$_1$: models at identical verdict accuracy differ by two orders of magnitude
	on triage noise, with only those maintaining baselines below 0.07 remaining
	operationally viable.
	Third, victim attribution in the vishing scenario is
	the specific behavioral capability that separates Tier~A from Tier~B:
	Devstral and Opus correctly exonerate Chris by reasoning that simultaneous
	anomalous and normal sessions from the same account identify a compromised
	credential rather than a malicious actor;
	all Tier~B models detect the attack
	but misclassify the victim. Fourth, a two-signal escalation threshold is
	architecturally exclusionary for single-surface threat profiles, demonstrating
	that SOC escalation parameters must be threat-class-specific.
	A prompt sensitivity analysis reveals that unstructured prompts induce
	vocabulary hallucination across model tiers --- semantically accurate but
	lexically non-canonical behavior labels that score as zero under exact-match
	evaluation.
	This effect is not limited to smaller models: both Tier~A models
	hallucinate vocabulary under \texttt{v2\_natural}, with Claude~Opus~4.6
	producing fully narrative behavior descriptions that the scorer cannot credit,
	including a vishing detection the scorer marks as absent.
	Verdict F$_1$ is
	stable across all prompt variants for all tested models, confirming that the
	capability gap between Tier~A and Tier~B reflects genuine reasoning differences
	rather than prompt calibration.
	The key distinction is that Tier~A models
	preserve their exoneration logic regardless of label vocabulary;
	Tier~B models
	lose the vishing detection path when taxonomy scaffolding is removed.
	
	OrgForge-IT is open source, configurable, and extensible.
	The SIEM format
	export enables direct practitioner use without custom parsers. The
	open-source evaluation harness enables community leaderboard contributions
	without re-running the simulation.
	The combination of verifiable ground
	truth, calibrated detection difficulty, and a completed multi-model
	leaderboard provides a foundation for systematic evaluation of LLM-based
	insider threat detection as the field matures.
	\section*{Acknowledgements}
	
	This work was conducted independently without external funding or
	institutional support.
	Benchmark design decisions were informed by the CERT
	dataset documentation, the Chimera evaluation framework~\citep{yu2025chimera},
	and scoring methodology from CyberSOCEval~\citep{cybersoceval2025}.

\end{document}